\documentclass[11pt]{article}
\pdfoutput=1
\usepackage{jheppub}
\usepackage{amsmath,amssymb,amsfonts,graphicx,slashed,color,amsthm,mathtools,upgreek, enumerate, tensor, subfig,braket}
\usepackage{arydshln}
\allowdisplaybreaks

\def\ben{\begin{equation}}
\def\een{\end{equation}}

  \let\g=\gamma  \let\e=\varepsilon
   
\let\l=\lambda     \let\r=v
 \let\t=\tau

 \let\G=\Gamma \let\D=\Delta

\let\pa=\partial
\def\be{\begin{equation}}
\def\ee{\end{equation}}
\def\ba{\begin{array}}
\def\ea{\end{array}}

\def\dalemb#1#2{{\vbox{\hrule height .#2pt
       \hbox{\vrule width.#2pt height#1pt \kern#1pt
               \vrule width.#2pt}
       \hrule height.#2pt}}}

\newcommand{\bea}{\begin{eqnarray}}
\newcommand{\eea}{\end{eqnarray}}

\let\tilde=\widetilde

\newcommand{\ttb}{T\overline{T}}

\title{Building Tensor Networks for Holographic States}
\author[a]{Pawel Caputa}
\author[b]{\!, Jorrit Kruthoff}
\author[b]{\!, Onkar Parrikar}

\affiliation[a]{Faculty of Physics, University of Warsaw, ul. Pasteura 5, 02-093 Warsaw, Poland} 
\affiliation[b]{Stanford Institute for Theoretical Physics, Department of Physics, Stanford University, CA 94305, USA.}

\abstract{We discuss a one-parameter family of states in two-dimensional holographic conformal field theories which are constructed via the Euclidean path integral of an effective theory on a family of hyperbolic slices in the dual bulk geometry. The effective theory in question is the CFT flowed under a $T\overline{T}$ deformation, which ``folds'' the boundary CFT towards the bulk time-reflection symmetric slice. We propose that these novel Euclidean path integral states in the CFT can be interpreted as continuous tensor network (CTN) states. We argue that these CTN states satisfy a Ryu-Takayanagi-like minimal area upper bound on the entanglement entropies of boundary intervals, with the coefficient being equal to $\frac{1}{4G_N}$; the CTN corresponding to the bulk time-reflection symmetric slice saturates this bound. We also argue that the original state in the CFT can be written as a superposition of such CTN states, with the corresponding wavefunction being the bulk Hartle-Hawking wavefunction. }

\date{September 2019}

\definecolor{webred}{rgb}{0.5, 0, 0}

\newcommand{\beq}{\begin{equation}}
\newcommand{\eeq}{\end{equation}}
\newcommand{\beqn}{\begin{eqnarray}}
\newcommand{\eeqn}{\end{eqnarray}}

\begin{document}

\maketitle
\parskip=12pt
\section{Introduction}

The Ryu-Takayanagi formula \cite{Ryu:2006bv} for the entanglement entropy of boundary subregions in AdS/CFT,
\beq
S_{EE} = \frac{1}{4G_N} \text{min}_{\gamma} \text{Area}(\gamma),
\eeq
is a powerful yet mysterious statement. It is powerful by virtue of the seemingly disparate subjects in theoretical physics (namely, quantum information theory and general relativity) which it brings together in one simple equation. But at the same time, it is mysterious, because we do not fully understand the deep connections it is alluding to. Of course, the RT formula (and to some extent its various generalizations \cite{Hubeny:2007xt, Faulkner:2013ana, Engelhardt:2014gca}) can be derived from the Euclidean path integral of gravity \cite{Lewkowycz:2013nqa}, but the derivation does not shed much light on the mystery of why quantum entanglement and geometry are so intimately tied together \cite{VanRaamsdonk:2010pw}. More recently, ideas coming from quantum error correction  \cite{Almheiri:2014lwa, Dong:2016eik, Harlow:2016vwg, Kang:2018xqy, Faulkner:2020hzi} have shed some light on this; in particular, it has been argued that the general structure of complementary recovery in quantum error correcting codes  fits nicely (albeit approximately) with AdS/CFT, and an ``RT-like formula'' for the entanglement entropy is a natural consequence of this general structure.\footnote{By ``RT-like formula'' we mean an expression for the entropy in terms of a bulk operator which lies in the center of the bulk algebra. See also \cite{Lewkowycz:2018sgn}, where it was argued that the precise RT formula for the entanglement entropy, i.e., $S_{EE} = A/4G_N$, follows as a general consequence of equality of bulk and boundary modular flows in AdS/CFT, which, the equality of bulk and boundary modular flows \cite{Jafferis:2015del}, in turn, follows on general grounds from the structure of complementary recovery in quantum error correcting codes \cite{Dong:2016eik, Harlow:2016vwg, Kang:2018xqy, Faulkner:2020hzi}.}

Tensor networks have played an important role in guiding our intuition about the structure of AdS/CFT \cite{Swingle:2009bg, Swingle:2012wq, Nozaki:2012zj, Haegeman:2011uy, Hartman:2013qma, Miyaji:2015yva, Miyaji:2015fia, Pastawski:2015qua, Yang:2015uoa, Hayden:2016cfa, Bao:2015uaa, Bao:2018pvs, Czech:2015kbp, Czech:2015xna, Miyaji:2016mxg, Bhattacharyya:2016hbx, Caputa:2017urj, Caputa:2017yrh, Boruch:2020wax, Milsted:2018san, Milsted:2018yur, Hu:2018hyd}, and in particular, the RT formula. Broadly speaking, a tensor network is an efficient way to construct a big tensor (say, the wavefunction of a state on a multi-partite Hilbert space) in terms of smaller tensors. In the context of AdS/CFT, a tensor network is a circuit representation for a quantum state in the CFT, which usually makes the entanglement structure of the quantum state manifest. For example, a tensor
network such as the multi-scale entanglement renormalization ansatz (MERA) \cite{Vidal:2007hda} is a variational ansatz for the wavefunction of the CFT state, which makes key use of the entanglement structure of the state from a position-space renormalization group perspective. In particular, the wavefunction is built as a quantum circuit, with successive layers of local operations called “disentanglers” and “isometries”. The basic idea is that starting from the UV state, at every layer of the circuit the disentanglers remove entanglement in the wavefunction at a given length scale, while the isometries coarse-grain and redefine the effective degrees of freedom relevant at the lower energy scale. This process is repeated scale by scale, until in the end we are left with a pure product state with no entanglement. This “emergent geometry” associated with the tensor network is clearly reminiscent of the bulk geometry in AdS/CFT, as has been discussed in \cite{Swingle:2009bg, Swingle:2012wq, Nozaki:2012zj, Haegeman:2011uy, Hartman:2013qma, Miyaji:2015yva, Miyaji:2015fia, Pastawski:2015qua, Yang:2015uoa, Hayden:2016cfa, Bao:2015uaa, Bao:2018pvs, Czech:2015kbp, Czech:2015xna, Miyaji:2016mxg, Bhattacharyya:2016hbx, Caputa:2017urj, Caputa:2017yrh, Boruch:2020wax, Milsted:2018san, Milsted:2018yur, Hu:2018hyd}. A key feature about such tensor networks is that they satisfy ``minimal-area'' bounds on the entanglement entropies of intervals in the boundary CFT, where by minimal-area we mean the minimum number of bonds/links in the network which we must cut in order to dissociate the boundary interval degrees of freedom from the rest of the state:
\beq
S_{EE}\leq (\log J) \times \text{min\;number\;of\;cuts}.
\eeq
The coefficient in the bound is the log of the bond dimension $J$, i.e., the local Hilbert space dimension of the severed links. Such a bound is reminiscent of the Ryu-Takayanagi formula for holographic entanglement entropy, with the key distinction being, of course, that the RT formula is an equality, not simply a bound. But this latter shortcoming can also be overcome; indeed, several models of tensor networks have been constructed which satisfy an RT-like \emph{equality} (see for instance, \cite{Pastawski:2015qua, Hayden:2016cfa}). However, to our knowledge, most of these constructions have remained within the realm of toy models, and an explicit tensor network construction for holographic CFT states is still lacking (although see \cite{Bao:2018pvs} for some recent progress). Another feature of these known tensor networks is that they prepare states with \emph{flat} entanglement spectra (i.e., where all the R\'enyi entropies are equal), which is clearly not the case for, say, the boundary CFT vacuum. This was recently clarified in \cite{Dong:2018seb, Akers:2018fow}, where the authors introduced the idea of ``fixed-area'' states to capture this behavior of tensor networks. Boundary CFT states are then not directly tensor networks, but rather superpositions there-of. 

\begin{figure}[t]
    \centering
    \includegraphics[width=0.4\textwidth]{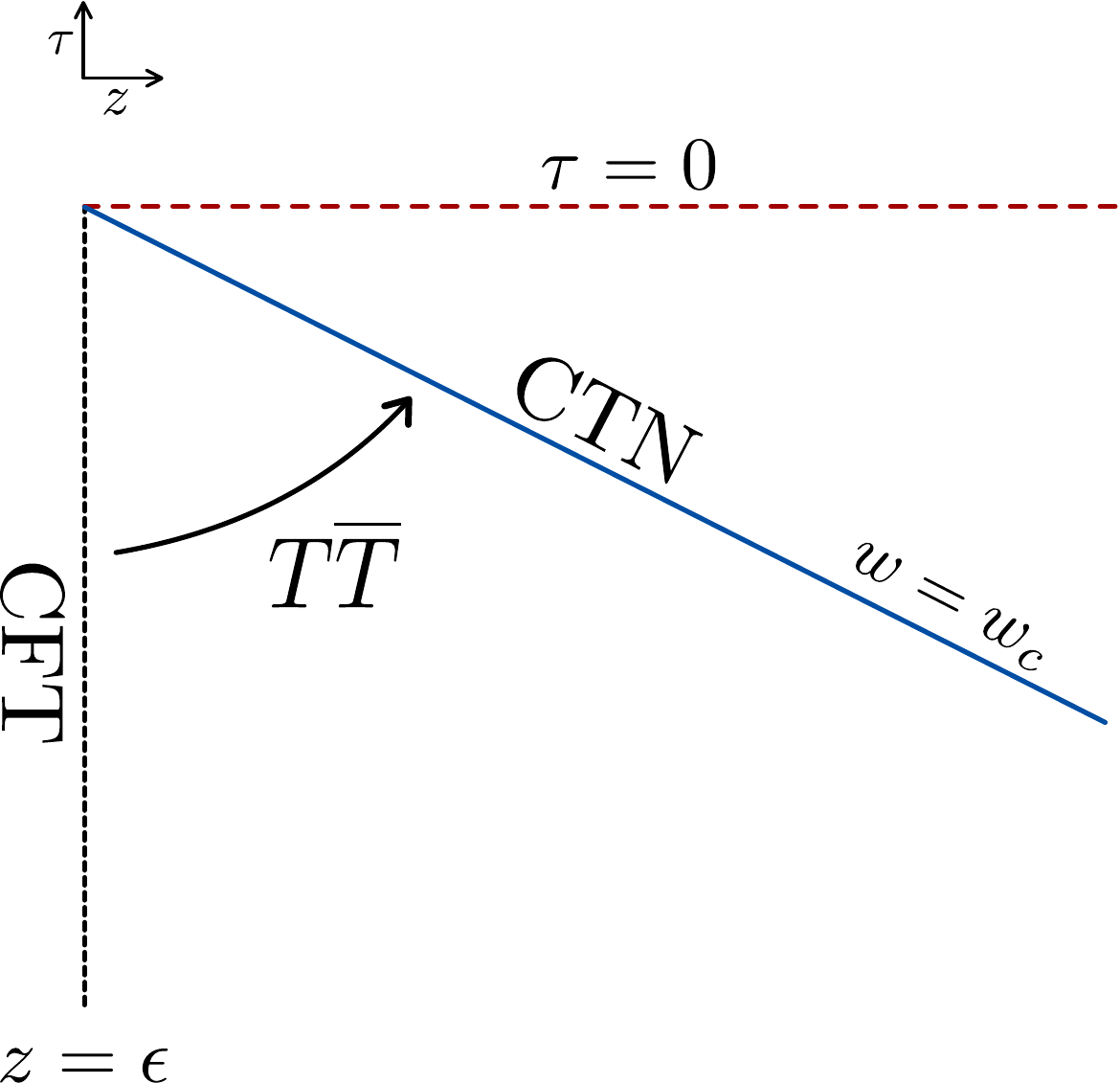}
    \caption{Summary of our construction. We consider the Euclidean path integral of a CFT on the lower half plane (black dotted line). We deform this path integral with a specific $T\bar{T}$ deformation, which is holographically dual to folding the slices (parametrized by $w = w_c$, blue solid line) into the bulk. The metric on the $w=w_c$ slices is that of $\mathbb{H}_2$ and the $T\bar{T}$ coupling $\l$ is related to the bulk coordinate $w_c$ as $\l = 2\pi G_N w_c^2$. Euclidean path integrals of the deformed theory on these slices are interpreted as continuous tensor network (CTN) states. The endpoint of the flow is at $w_c = 1$, where the ``folded" slice coincides with the time-reflection symmetric slice in the bulk at $\tau = 0$ (red dashed line).}
    \label{fig:Summary}
\end{figure}

The goal of this work is to propose a concrete way to formalize the AdS/tensor network correspondence. Our main tool will be the $\ttb$ deformation of the boundary conformal field theory, and the conjecture of McGough, Mezei and Verlinde \cite{McGough:2016lol} that the boundary $\ttb$ deformation is dual to the bulk gravitational theory with a finite radial cutoff. In the standard radial setup, one can imagine turning up the $\ttb$ coupling as evolving the boundary state ``radially'' inwards. This radial evolution was studied in \cite{Kruthoff:2020hsi}, where it was argued that it generates a superposition of tensor networks. Here, we will take a slightly different approach inspired by the discussion in \cite{Belin:2020oib} and consider a $\ttb$ deformation which ``folds in'' the Euclidean path-integral of the boundary CFT towards the time-reflection symmetric slice in the bulk geometry. More precisely, we consider the one-parameter family (where the parameter will be labelled $w_c$) of lower half planes in Euclidean signature interpolating between the lower half-space in the asymptotic boundary and the bulk time-reflection symmetric slice. The metric on each slice is that of hyperbolic space with a radius of curvature which depends on $w_c$. Our main proposal is that we should interpret the states obtained from the Euclidean path integral of the $\ttb$-deformed field theory (with coupling $\l= 2\pi G_N w_c^2$ on the half-plane at $w_c$) as a family of \emph{continuous tensor network} (CTN) states (see figure \ref{fig:Summary}, and also see figure \ref{fig:uv/ir} for more detail).

We can interpret a Euclidean path integral as a continuous tensor network as long as it is built from Euclidean evolution with respect to a Hamiltonian with some notion of locality; the scale of locality, which in the present case is set by $\sqrt{\l}$, can be interpreted as the ``size'' of the constituent tensors. One piece of evidence for our proposal is that the above CTN states satisfy an upper bound on the entanglement entropy of boundary intervals in terms of a ``minimal-area'' cut through the path-integral, with the crucial property that the log of the bond dimension is $\frac{1}{4G_N}$. In the limit where the folded slice approaches the time-reflection symmetric slice in the bulk, the bound saturates to the Ryu-Takayanagi formula. Our construction also clarifies the distinction between the boundary CFT state and the CTN states -- although they share the same large $N$ entanglement entropies, the boundary CFT state can be shown to be a superposition of the CTN states, with the wavefunction being the Hartle-Hawking wavefunction.

As we already reviewed, the interpretation of slices of the dual geometry as tensor networks dates back to the very beginning of the subject \cite{Swingle:2009bg}. Indeed, several different perspectives on this AdS/tensor network correspondence have already appeared in the literature. Let us contrast our $T\bar{T}$ proposal with two recent constructions based on path integrals in CFT: firstly, in \cite{Milsted:2018san,Milsted:2018yur} the authors demonstrated that CFT path integrals on constant (negative, positive or zero) curvature slices of, Euclidean or Lorentzian, $AdS$ spacetimes represent particular evolution's of wave functions with CFT generators which can be interpreted as tensor networks (for e.g., the MERA network corresponds to the path integral on a light-sheet). The main difference between these works and ours is that our theory on the hyperbolic slice is not a CFT but rather a particular regulated, effective theory, i.e., the $T\bar{T}$ deformed CFT. 
Another concrete approach to holographic tensor networks is path integral optimization \cite{Caputa:2017yrh,Caputa:2017urj}. In this approach one finds the optimal metric for Euclidean path integrals by minimizing a certain functional (path integral complexity) and also finds hyperbolic slices of AdS geometries dual to CFT states prepared by the path integral.\footnote{See also \cite{Takayanagi:2018pml} for generalization to arbitrary slices of holographic geometries.} The path integral optimization procedure was recently interpreted using standard tools of AdS/CFT as maximization of the Hartle-Hawking wave functions in the bulk that reproduces the same optimal metrics \cite{Boruch:2020wax}. Moreover, the construction was generalized to a one-parameter family of hyperbolic slices (with constant mean curvature) in the bulk, interpreted as unoptimized tensor networks, which are the same as we discuss here. On the field theory side, the advantage of $T\bar{T}$ is that it provides a precise interpretation of the bulk slices as holographic tensor networks, in terms of a universal deformation of the CFT. From the gravity perspective, the main difference is that our $T\bar{T}$-based construction corresponds to Dirichlet boundary conditions on the ``folded'' cut-off slice, while that of \cite{Caputa:2017yrh,Caputa:2017urj} corresponds to Neumann boundary conditions, standard in AdS/BCFT \cite{Takayanagi:2011zk}. While the slices we choose here are constant mean curvature slices, we expect that our arguments generalize to a more general class of bulk slices.\footnote{In a sense, we, as well as \cite{Milsted:2018san,Milsted:2018yur}, should supplement our prescription with a minimizaion/maximization of some notion of ``complexity'' which singles out these slices. See also the discussion section.} We hope that further investigation into both approaches will shed more light on the precise differences and similarities between these clearly related ideas.

While our proposal provides a potentially concrete tool to formalize the AdS/tensor network correspondence, we should however mention that the subject of $\ttb$ deformations is not fully settled. In particular, the $\ttb$ operator is irrelevant, and at least with the holographic sign, naively leads to a complexification of the energy spectrum beyond the energy $\frac{1}{4\lambda}$ (at zero spin), where $\lambda$ is the $\ttb$ coupling. This is an indication that the deformation drastically modifies the UV limit of the theory and good UV completions are not known. Thus, in a sense, our results transmute the problem of understanding the AdS/tensor network correspondence to the problem of resolving the UV structure of $\ttb$ deformed CFTs. Furthermore, the precise definition of the $\ttb$ operator at finite $N$ is not understood on general curved backgrounds (see \cite{Jiang:2019tcq, Brennan:2020dkw} for further discussion on this issue), although at large $N$, which is the limit we are interested in, we can use large $N$ factorisation to define it. Nevertheless, we feel that it is worthwhile establishing this connection between the $\ttb$ deformation and tensor networks, in the hope that it will shed some light on the entanglement structure of holographic states and its geometric realization in AdS/CFT.

The rest of the paper is organized as follows: in section \ref{sec:tn}, we will begin with a brief review of the $\ttb$ deformation. We will then spell out the details of the ``folding'' deformation and explain its interpretation as generating a one-parameter family of continuous tensor network states. In section \ref{sec:ee}, we will argue that these tensor networks satisfy a minimal-area bound on the entropy of boundary intervals, and that in the limit where we approach the time-reflection symmetric slice in the bulk, the bound saturates to the Ryu-Takayanagi formula. We will discuss some open questions and future directions in the discussion section.

\section{Tensor networks for holographic states}\label{sec:tn}
In this section, we discuss our proposal for a tensor network interpretation of holographic states. Since our construction will involve a $T\overline{T}$ deformation of the boundary CFT, we will begin with a lightning review of this subject.

\subsection{Brief review of $T\overline{T}$}

The $T\overline{T}$ deformation is an irrelevant deformation of two-dimensional quantum field theories that is quadratic in the stress tensor, first studied by Zamolodchikov \cite{Zamolodchikov:2004ce, Smirnov:2016lqw}. Typically it is written in the form of a flow equation of the action $S$ as
\be\label{floweq}
\partial_\l S[\Phi;\l] = - \int d^2 x \sqrt{\gamma} \,\e^{\mu\nu}\e^{\rho\sigma} T^{(\l)}_{\mu\rho}(x)T^{(\l)}_{\nu\sigma}(x),
\ee
where $\Phi$ denotes a collection of the elementary fields the theory is written in terms of, $\l$ is the coupling constant with dimensions of length squared and $T^{(\l)}_{\mu\nu}(x)$ is the stress tensor of the theory at coupling $\l$. Thus, at a given value of $\l$, we deform by the operator on the RHS of equation \eqref{floweq} -- often called the $\ttb$ operator -- built out of the stress tensor at that value of $\l$, giving a step-by-step definition of the deformation as a ``flow''. Despite the irrelevance of the $\ttb$ operator, we can still compute various quantities non-perturbatively in $\l$, such as the deformed energy levels \cite{Smirnov:2016lqw}, partition function and S-matrix \cite{Freidel:2008sh, Cavaglia:2016oda, Cardy:2018sdv, Dubovsky:2018bmo, Dubovsky:2017cnj, Datta:2018thy, Caputa:2019pam,Caputa:2020lpa, Tolley:2019nmm, Mazenc:2019cfg}, entanglement entropy \cite{Donnelly:2018bef, Lewkowycz:2019xse}, correlation functions \cite{Kraus:2018xrn, Aharony:2018bad, Cardy:2019qao, He:2019vzf}, energy eigenstates \cite{Kruthoff:2020hsi} etc. One of the basic reasons behind this calculability is the factorization of the $\ttb$ operator on the plane and cylinder. For e.g., the energy of a state with angular momentum $j$ in the deformed theory on a circle of length $L$ can written in terms of the undeformed energy as \cite{Smirnov:2016lqw}
\beq\label{EFlow}
E(\l) = \frac{L}{4\l}\left(1 - \sqrt{1-\frac{8\l E(0)}{L}+\frac{16j^2\l^2}{L^2}}\right).
\eeq
Note that this energy becomes complex at some point. With $j=0$ for instance, this happens at $E(0) = \frac{L}{8\l}$. We may interpret this as a cutoff on the theory.

For our purposes, however, another facet of the $T\overline{T}$ deformation will be relevant. In \cite{McGough:2016lol} it was conjectured that for a holographic conformal field theory dual to pure gravity, the $\ttb$ deformed theory is dual to gravity in AdS with Dirichlet boundary conditions at some \emph{finite radius} $r_c$. If we choose the Fefferman-Graham gauge for the bulk metric\footnote{We will work in units where $\ell_{\rm AdS} = 1$.},
\beq\label{FG}
ds^2 = \frac{dr^2}{r^2} + g^{(0)}_{\mu\nu}(r,x) dx^\mu dx^\nu,\;\;\;\; g^{(0)}_{\mu\nu}(r,x) \equiv r^2 \gamma_{\mu\nu}(r,x),
\eeq
then the relation between the $T\overline{T}$ deformation and holography with a finite-radius cutoff is easily seen by noting that evolution of the boundary partition function in the radial direction in AdS$_3$ is governed by a radial Wheeler-DeWitt equation, i.e., the radial Hamilton constraint
\beq\label{radialH}
K^2 - K_{\mu\nu}K^{\mu\nu} - R_{g^{(0)}}- 2 = 0,
\eeq
where $K_{\mu\nu}$ is the extrinsic curvature of the slice $r= r_c$, and $R_{g^{(0)}}$ is the Ricci scalar of the induced metric $g^{(0)}$. Using the Balasubramanian-Kraus formula \cite{Balasubramanian:1999re} to convert the extrinsic curvature into the boundary stress tensor
\beq\label{TBK}
T_{\mu\nu} = \frac{1}{8\pi G_N}\left(K_{\mu\nu}- K g^{(0)}_{\mu\nu} + g^{(0)}_{\mu\nu}\right),
\eeq
and writing the induced metric in terms of the field theory background metric $g^{(0)}_{\mu\nu} = r_c^2 \gamma_{\mu\nu}$, the constraint takes the form,
\be \label{TFE}
{T^{\mu}}_{\mu} = -\frac{4\pi G_N}{r_c^2}\left(T_{\mu\nu}T^{\mu\nu}-({T^{\mu}}_{\mu})^2\right) -\frac{1}{16\pi G_N} R_{\gamma},
\ee
where note that now all indices are raised/lowered/contracted with the metric $\gamma_{\mu\nu}$, and $R_{\gamma}$ is the Ricci scalar of $\gamma_{\mu\nu}$. This is often referred to as the \emph{trace flow equation}. In order to translate this to a flow equation for the boundary field theory action, we note that assuming there is only one non-trivial scale in the theory generated by a dimensionful coupling $\l$ (which is the case provided the seed theory is a CFT), then \footnote{Here $\pa_{\l}$ denotes a partial derivative with respect to $\l$ with the background metric $\gamma$ fixed.}
\beq
\left(\Delta_{\l}\l\pa_{\l}  +  \int d^2x\,2\gamma^{\mu\nu}\frac{\delta}{\delta \gamma^{\mu\nu}}\right) \log\,Z = 0 \quad \Rightarrow \quad  \l\pa_{\l} S_{\text{eff}} =  \frac{1}{\Delta_{\l}}\int d^2x\,\langle {T^{\mu}}_{\mu}\rangle,
\eeq
with $\D_{\lambda} = -2$. Therefore, using equation \eqref{TFE}, we find that the effective action of the boundary theory flows under the $\ttb$ deformation, with the coupling constant given by
\be
\l = \frac{2\pi G_N}{r_c^2},
\ee
with $r_c$ the radial location of the boundary. This procedure of finding the appropriate deformation of the boundary theory that corresponds to Dirichlet boundary conditions at finite $r$ also holds in other dimensions \cite{10.21468/SciPostPhys.9.2.023, Gross:2019uxi, Gross:2019ach, Hartman:2018tkw, Taylor:2018xcy,Caputa:2019pam, Belin:2020oib}. However in more than two boundary dimensions, the aforementioned factorisation property only holds at large $N$. 
Also note that $\g_{\mu\nu}$ is not restricted to be $r$ independent; the above argument also holds, for instance, in black hole geometries. Given the conjectured relation between the $\ttb$ deformation and gravity with a radial cut-off, it seems natural that the flow of, say, the vacuum state under the $\ttb$ deformation would generate a bulk tensor network. Indeed, it was shown in \cite{Kruthoff:2020hsi} that the $\ttb$ flow in fact generates a superposition of tensor networks. Here, we will try to make this precise from a slightly different perspective, inspired by the discussion in \cite{Belin:2020oib}.

\subsection{Folding the boundary in}
We now move on to our main topic of interest, namely using $T\overline{T}$ to formalize the AdS/tensor network correspondence. We start with the boundary CFT in the vacuum state $|0\rangle$ on the real line, and consider either the overlap $Z_{CFT} = \langle 0 | 0\rangle$, or some correlation function of operators inserted at $\tau=0$ in the CFT. Either way, the object of interest in the boundary CFT  is the Euclidean path-integral over the entire Euclidean plane. The discussion here can be generalized to the case where stress tensor sources are turned on in the Euclidean path integral or even to the thermal state, as long as we maintain time reflection symmetry $\tau \to - \tau$ (where $\tau$ is Euclidean time); for simplicity of presentation, we will restrict to the vacuum state in our discussion. We will further assume that the CFT is holographic, with the bulk dual consisting purely of Einstein gravity. Whether such a bulk theory can be dual to a single CFT or requires an ensemble interpretation has not been settled yet (see, for instance, \cite{Cotler:2020ugk} and references therein). However, these considerations are beyond the scope of the present paper; here we will simply assume that we can take the bulk dual to be Einstein gravity, and limit ourselves to semi-classical (i.e., perturbative in $G_N$) considerations. 

Coming back to the object of interest, $Z_{CFT}$, using the standard AdS/CFT dictionary, this can be written as a bulk path-integral
\beq\label{CFTPI}
Z_{CFT} = \int_{g^{(0)}=e^{2\Omega}\delta} Dg \;e^{-S_{\rm bulk}[g]}, 
\eeq
where $S_{\rm bulk}$ is the Einstein-Hilbert action with a cosmological constant $\Lambda = -1$, plus the Gibbons-Hawking boundary term at asymptotic infinity, as well as the standard counterterm. The subscript indicates that we must fix Dirichlet boundary conditions on the metric at asymptotic infinity, with the induced metric there restricted to be conformally flat. In the large $N$ limit ($G_N \to 0$) , the gravitational path integral is dominated by a classical solution, namely $AdS_3$
\beq
g = \frac{dz^2+ d\tau^2+dx^2}{z^2}.
\eeq
We can view the semi-classical path integral as an integral over small metric fluctuations around $AdS_3$ with Dirichlet boundary conditions at the asymptotic boundary
\beq
g_{\mu\nu}(\epsilon, x)= \frac{1}{\epsilon^2}\delta_{\mu\nu},
\eeq
where putting the asymptotic boundary at $z=\epsilon$ is a standard way to introduce a UV cut-off in AdS/CFT. Our goal now is to give a tensor network interpretation to the time-reflection symmetric slice in the bulk saddle point geometry. We will often denote the metric on this slice as $\gamma^{(\star)}$:
\beq
\gamma^{(\star)} = \frac{dz^2 + dx^2}{z^2}.
\eeq
In fact, what we will find is a one-parameter family of tensor network states which interpolates between the geometry of the lower-half Euclidean space in the asymptotic boundary and the time-reflection symmetric slice in the bulk.

\begin{figure}
    \centering
  \begin{tabular}{c c c}  \includegraphics[height=4.4cm]{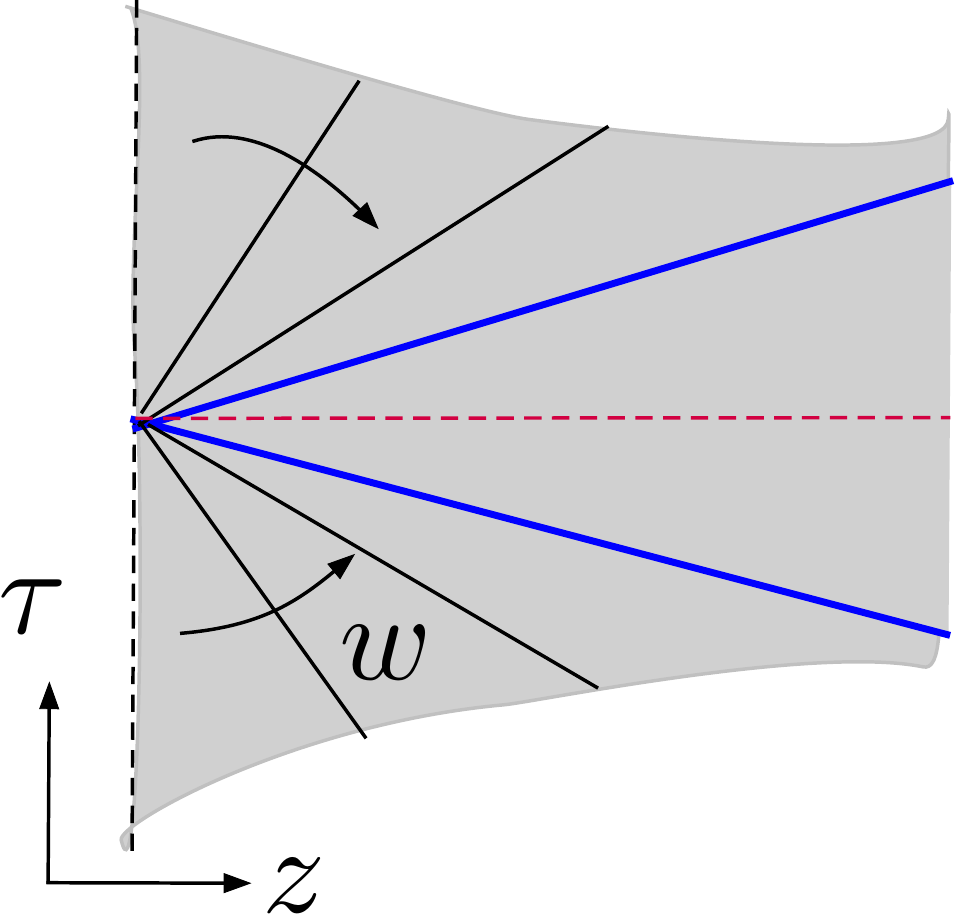} & \hspace{2cm} &  \includegraphics[height=4.2cm]{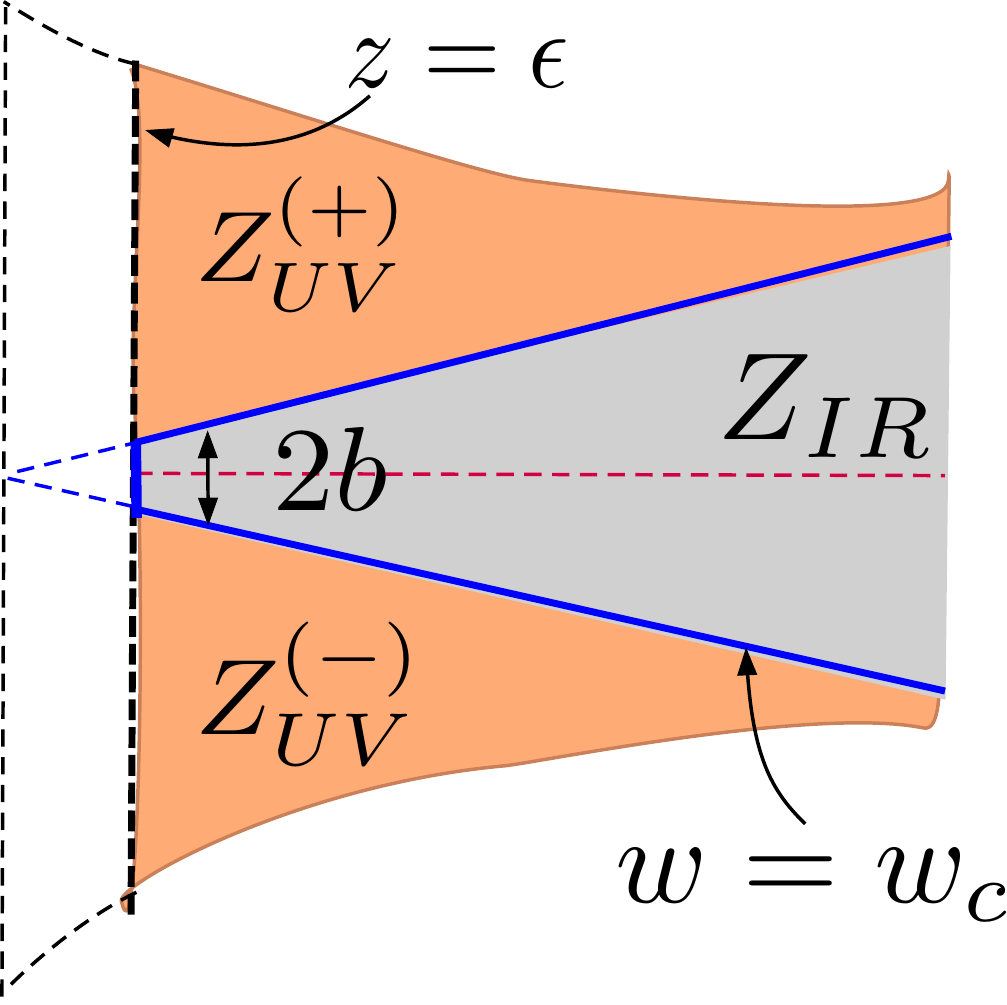}
  \end{tabular}
    \caption{\textbf{Left}: Folded slices parametrized by $w$. The red dashed line is the time-reflection symmetric slice. \textbf{Right}: Same blue slice as in the left panel, but regulated with a UV cutoff at $z=\epsilon$; this leads to a flat space regulator strip in the boundary of width $2b$. $Z_{UV}^{(\pm)}$ are the gravity path integrals over the upper and lower orange regions, whereas $Z_{IR}$ is the gravity path integral over the wedge (grey). }
    \label{fig:uv/ir}
\end{figure}

To this end, let us consider a new coordinate system for $AdS_3$, where we essentially switch to polar coordinates in the $(\tau, z)$ plane
\beq
z= u \cos\theta,\qquad \tau = u \sin \theta.
\eeq
In these coordinates, the AdS metric becomes
\beq
g = \frac{d\theta^2}{\cos^2\theta}+\frac{1}{\cos^2\theta}\frac{du^2+dx^2}{u^2}.
\eeq
Introducing a new coordinate $w$ defined by $\cos\theta = \frac{2w}{1+w^2}$,
we can rewrite the metric as
\beq
g = \frac{dw^2}{w^2}+ \frac{f(w)}{w^2}\frac{(du^2+dx^2)}{u^2},\;\;\;\;\;\;f(w)= \frac{1}{4}(1+w^2)^2,
\eeq
where $w\in [0,1]$ and we can now let the coordinate $u$ run from $-\infty$ to $\infty$ with the sign of $u$ corresponding to the sign of $\tau$. Note that constant $w$ slices are essentially ``folded'' images of the boundary, where we can think of the coordinate $u$ as the new Euclidean time coordinate. As $w\to 0$, these constant $w$ slices approach the asymptotic boundary, while as $w \to 1$, these  slices ``fold in'' towards the time-reflection symmetric slice in the bulk (see the left panel of figure \ref{fig:uv/ir}). 

The basic idea now is to regard the coordinate $w$ as the new radial coordinate. As reviewed in the previous subsection, putting the boundary at finite radial slices in the bulk has been argued to be dual to a $T\overline{T}$ deformation of the boundary CFT. In the standard $\ttb$ literature, this ``radial'' coordinate is taken to be the Poincare coordinate $z$, but the discussion in the previous section was valid for a general metric of the form in equation \eqref{FG}. Thus, the bulk gravity theory inside the ``wedge'' region $w>w_c$ should be dual to an effective boundary field theory on the slice $w=w_c$; following the logic of the previous section, the effective field theory is the $T\overline{T}$ deformation of the boundary CFT. To be more precise, if we denote by $g^{(0)}$ the induced metric on the slice $w=w_c$, then we can compute the \emph{semi-classical}, gravity path integral in the wedge $w>w_c$, with Dirichlet boundary conditions at $w=w_c$ (see the right panel of figure \ref{fig:uv/ir}):
\beq
g^{(0)}_{\mu\nu}(w_c,u,x) = \frac{1}{w^2_c}\gamma_{\mu\nu}(w_c,u,x).
\eeq
This bulk path-integral -- which we will call $Z_{IR}[w_c, \gamma]$ -- should then be equal to the Euclidean path-integral of an effective boundary field theory ``living'' at the slice $w=w_c$, with the background metric $\gamma_{\mu\nu}$. Following the discussion in the previous subsection, the Wheeler-DeWitt equation for $Z_{IR}$, together with the Balasubramanian-Kraus prescription for the boundary stress tensor imply that we can take the effective field theory on the slice $w_c$ to be the boundary CFT flowed under the $T\overline{T}$ deformation, living on the background metric $\gamma_{\mu\nu}$ and with the $T\overline{T}$ coupling given by $\lambda = 2\pi G_N w_c^2$ (see details in Appendix \ref{Details}). We should mention that a similar interpretation of the hyperbolic slicing of AdS in terms of $\ttb$ has also previously appeared in \cite{Gorbenko:2018oov}.\footnote{Our setup is also similar to that of wedge holography \cite{Akal:2020wfl}, but with wedge in Euclidean time $\tau$ instead of position $x$.} It is also important to emphasize again that we are restricting ourselves to the semi-classical limit, where we have a background AdS geometry and we are integrating over metric fluctuations around this background; in particular, we do not include the contribution of topology changing geometries inside the bulk path integral.

The crucial point now is that we have a Euclidean path-integral of an effective theory at the $w=w_c$ slice, living on a background metric $\gamma$. If we let $\gamma^{(+)}$ be the restriction of this background metric on the upper half plane, and $\gamma^{(-)}$ be the restriction to the lower half plane, then in the effective boundary field theory, the Euclidean path-integral has the interpretation of an overlap\footnote{ $\g^{(+)}$ is the time-reflection of $\g^{(-)}$ in the saddle geometry. However, in the more general setup relevant to equation \eqref{overlap} below, they will be different.}:
\beq
Z_{IR}[w_c, \gamma] = \langle w_c, \gamma^{(+)}  | w_c, \gamma^{(-)}\rangle_{T\overline{T}}, 
\eeq
where $|w_c, \gamma^{(-)}\rangle_{T\overline{T}}$ is a state defined by the lower-half Euclidean path-integral in the $T\overline{T}$ deformed theory, with the background metric $\gamma^{(-)}(w_c)$ and the $T\overline{T}$ coupling $\lambda = 2\pi G_N w_c^2$.\footnote{The notation $|w_c, \gamma^{(-)}\rangle$ is a bit redundant, because as discussed in \cite{Belin:2020oib}, the effective theory only depends on the ratio $\frac{\gamma_{\mu\nu}}{\l}$. } One important detail which we must address here is that in order to prevent UV divergences from the $u \to 0$ region, we must put a UV cut-off. We will do this by using the asymptotic cut-off slice at $z=\epsilon$, where the undeformed CFT is defined. When 
\beq
u_{0}= \pm \frac{(1+w_c^2)}{2w_c} \epsilon,
\eeq
the constant $w_c$ surface intersects with this asymptotic boundary cutoff surface $z=\epsilon$. The actual boundary surface will thus be specified by constant $w=w_c$ for $|u|>u_0$ (i.e., the folded upper and lower half planes), glued together with a ``regulator strip'' of flat space at constant $z=\epsilon$ (see the right panel of figure \ref{fig:uv/ir}). The thickness of this regulator strip is given by 
\beq
2b = \frac{(1-w_c^2)}{w_c}\epsilon.
\eeq
In the limit $w_c \to 1$, the regulator $b\to 0$. A concern here is that this regulator introduces a kink in the field theory path integral at $u=u_0$; this can be avoided by choosing a smooth boundary slice (i.e., with the kink at $u_0$ smoothed out). Alternatively, we could add the Hayward term at the corner, as was done in \cite{Akal:2020wfl}, but we expect these choices to only affect the very UV details of the entanglement structure of the state\footnote{Our setup is also similar to \cite{Chen:2020tes}, where a gravitational region on Euclidean AdS$_2$ coupled to a flat reservoir was considered. In our case, the ``gravity" region is replaced by the $T\bar{T}$ deformation and our reservoir is small (scaling with $\epsilon$). We don't expect there to be entanglement islands in this case though, because the $T\bar{T}$ deformation does not allow topology change.}.

The states $|w_c, \gamma^{(-)}\rangle_{\ttb}$ were discussed recently in \cite{Belin:2020oib} in the context of Hartle-Hawking wavefunctions in the bulk, where they were called ``metric eigenstates''. Our main message here is that we should interpret these states $|w_c, \gamma^{(-)}\rangle_{\ttb}$ as a one-parameter family of \emph{continuous tensor network} (CTN) states, with the parameter $0 < w_c< 1$. The tensor networks in question are not circuits built out of discrete elements, but ``continuous networks'' written in the form of Euclidean path-integrals (in the spirit of \cite{Milsted:2018yur} but for the $T\bar{T}$ deformed CFT). As discussed above, the path integral is performed on the background metric $\gamma^{(-)}(w_c)$, which in the AdS case is given by
\beq\label{metricCFT}
\gamma^{(-)}(w_c)= f(w_c) \frac{(du^2+dx^2)}{u^2}.
\eeq
Even though the state is defined via this Euclidean path-integral preparation using the $\ttb$ deformed field theory, we should think of this state as nevertheless living in the UV Hilbert space at $\tau = 0$ (and $z=\epsilon$) in the asymptotic boundary. Nevertheless, the geometry of the bulk slice $w=w_c$ is crucially encoded in this asymptotic state, via its preparation using the $\ttb$ path-integral. Note also that the $\ttb$ theory is expected to be non-local at the length scale 
\beq
\ell_{NL} \sim \sqrt{\l} \sim \frac{1}{\sqrt{c}}w_c.
\eeq
We may regard this as the size of the the individual tensors in the network. Interestingly, the scale of non-locality is way below the AdS scale, i.e., $\ell_{NL}\ll \ell_{\rm AdS}$.\footnote{Of course, the conformal factor of $u^2$ in the metric enhances $\ell_{NL}$ at large $u$ (see also below), but we can regulate this by putting an IR cut-off at large $u$ or at large $z$, similar to the UV cut-off.} 

\subsubsection*{Two alternative perspectives}

Besides the perspective we have given above, there are two other perspectives one can take on the path integral preparation of CTN states. The first one is rather simple and involves stripping off the factor $f(w)$ from the metric and putting it into $\l$; this amounts to what was called in \cite{Hartman:2018tkw} the ``total" flow of the effective field theory. In this perspective, the radius of curvature of the background hyperbolic space metric is fixed, as the total flow of $\l$ takes the would-be non-trivial flow of the metric into account. In our case this is easy to do because the would-be $w_c$ dependence of the metric is just a conformal factor out front. One can check, using the formulae in \cite{Hartman:2018tkw} that this results in a change in the identification between $\l$ and $w_c$. As expected one gets,
\be 
\l = \frac{2\pi G_N w_c^2}{f(w_c)} = \frac{8\pi G_N w_c^2}{(1+w_c^2)^2}.
\ee
This formula also tells us something interesting. As a function of $w_c$, $\l$ has a maximum value at $w_c = 1$ (this is due to the $w_c \to 1/w_c$ symmetry of this relation) and so not all values of $\l$ correspond to a slice in the bulk and in particular $0\leq \l \leq 2\pi G_N$ in order to correspond to real $w_c$. This makes it natural that the folding deformation can only be applied till some finite $\l$ for the holographic interpretation to uphold.

\begin{figure}
    \centering
    \includegraphics[height=3.5cm]{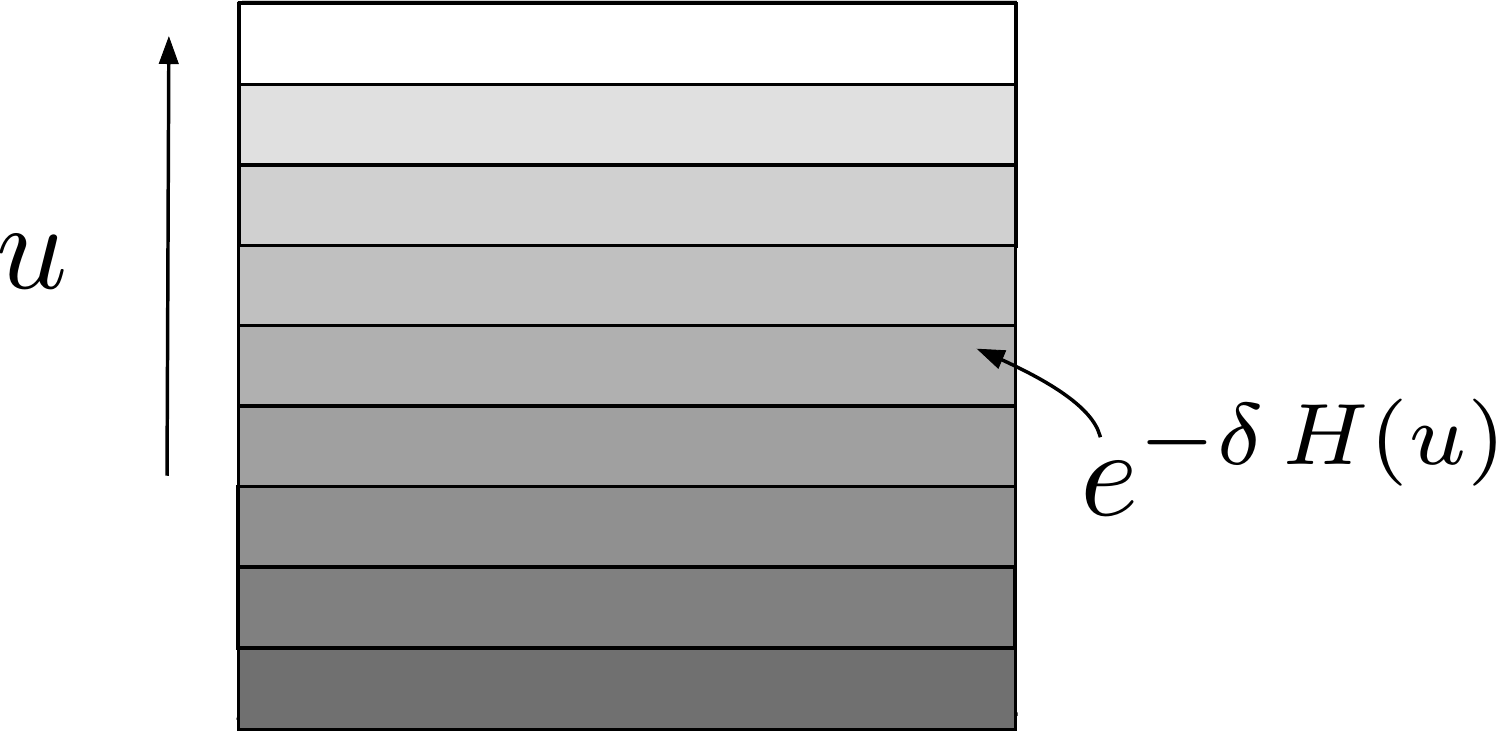}
    \caption{An alternative picture of the folding deformation. The CFT lives on flat space, but at each time step the $T\bar{T}$ coupling is decreased slightly, holographically this will result in a folded slice inside the bulk, since we go deeper in the bulk for earlier times $u$.}
    \label{fig:state}
\end{figure}

Yet another perspective on this state preparation (albeit schematic, but which we expect to make sense at large $N$) is to take the background metric to be flat and put the non-trivial $u$-dependent conformal factor into a position dependent $\ttb$ coupling: $\lambda \propto u^2$ \cite{Belin:2020oib}. From this perspective, the state is obtained by successive infinitesimal Euclidean time evolutions with weaker and weaker coupling (see figure \ref{fig:state}),\footnote{The $\ttb$ deformation on the plane in finite energy states can be shown to be a canonical/Bogoliubov transformation \cite{Kruthoff:2020hsi}. We thus expect the Hilbert spaces at different $u$ to be isomorphic on the plane. }
\beqn
|w_c, \gamma^{(-)}\rangle_{\ttb} &=& \lim_{\delta \to 0}\;e^{-bH_0}e^{-\delta\,H(-\delta)} e^{-\delta\,H(-2\delta)} e^{-\delta\,H(-3\delta)} \cdots |\psi_0\rangle,\nonumber\\
&=&e^{-bH_0}\mathcal{T}\exp\left( -\int_{-\infty}^{u_0} du'H(u')\right)|\psi_0\rangle,
\eeqn
where $H(u)$ is the Hamiltonian for the flat space theory at $\ttb$ coupling $\l = \frac{2\pi G_N w_c^2}{f(w_c)} u^2$, and $\psi_0$ is some initial, fiducial state. If we take this state to be a completely unentangled state, then we may view the above evolution as building the entanglement structure of $|w_c, \gamma^{(-)}\rangle_{\ttb}$ scale-by-scale, with the operator $e^{-\delta\; H(u)}$ adding entanglement at the length scale $u$.\footnote{This is somewhat reminiscent of the construction in \cite{Simidzija:2020ukv}.} From a holographic point of view, the time-dependence of the $T\bar{T}$ coupling is reflected by the fact that at successive time-step the boundary is pushed less and less inside the bulk.

\subsubsection*{The vacuum as a superposition of tensor networks}

We can now ask how the CTN states $|w_c,\gamma\rangle_{\ttb}$ are related to the vaccum state $|0\rangle_{CFT}$ in the boundary CFT, which is our primary object of interest. Going back to the path-integral \eqref{CFTPI} for $Z_{CFT} = \langle 0|0\rangle_{CFT}$, we can ``slice open'' the semi-classical path-integral of gravity on the slice $w=w_c$: 
\beq\label{split}
Z_{CFT} = \int Dg^{(0)}\; Z_{UV}[w_c,g^{(0)}]Z_{IR}[w_c,g^{(0)}],
\eeq
where $g^{(0)}=\frac{1}{w_c^2}\gamma$ is the induced metric on the surface $w=w_c$. $Z_{UV}$ also has Dirichlet boundary conditions at the asymptotic boundary, but these will be left implicit. Note that such a ``cutting open'' of the path integral is problematic in the full gravitational path-integral, with topology change etc., because the location of the surface at which to cut open is hard to specify given that bulk topology is being summed over and in addition the metric is being integrated over. However, as we have emphasized previously, at the level of the semi-classical path integral where we integrate over metric fluctuations around the background AdS geometry, this procedure seems well-defined. At any rate, a version of equation \eqref{split} can be derived explicitly using the Hubbard-Stratonovich trick to simplify the $\ttb$ deformation; see, for instance, \cite{Belin:2020oib}. With our previous interpretation of $Z_{IR}$ in terms of an overlap in the $\ttb$ deformed field theory, we can now write
\beq\label{overlap}
\langle 0|0\rangle_{CFT} = \int Dg^{(0)}_+Dg^{(0)}_-\; Z^{(+)}_{UV}[w_c,g^{(0)}_+]Z^{(-)}_{UV}[w_c,g^{(0)}_-]\langle w_c, \gamma^{(+)}  | w_c, \gamma^{(-)}\rangle_{T\overline{T}},
\eeq
where we have split the integration over $g^{(0)}$ and the UV part of the gravity path integral into the upper half portion ($\tau >0$) labelled by the plus sign, and the lower half portion $(\tau < 0)$ labelled by the minus sign (see the right panel of figure \ref{fig:uv/ir}).  While we have written this equation for the norm of the CFT vacuum state, it is equally valid for correlation functions of a small number of stress tensor insertions in the boundary CFT at $\tau=0$. This therefore suggests the identification
\beq \label{tn1}
|0\rangle_{CFT} \sim \int Dg^{(0)}_-\,Z^{(-)}_{UV}[w_c,g^{(0)}_-]\; |w_c,\gamma^{(-)}\rangle_{T\overline{T}},
\eeq
where again $Z^{(-)}_{UV}$ is the lower-half portion of the UV gravity path-integral, i.e., in the region $\tau < 0$ and $w<w_c$.\footnote{The flat regulator strip in the boundary separating the upper and lower half-planes is crucial here, so that we can cleanly separate the gravitational path integrals.} Note that equation \eqref{tn1} is a formal restatement of \eqref{overlap}, which is only valid for evaluation of correlation functions with a small number (not scaling with $N$) of stress tensor insertions at $\tau=0$. Nevertheless, we may interpret equation \eqref{tn1} as giving an approximate representation of the CFT vacuum state in terms of a \emph{superposition} of the CTN states $|w_c,\gamma^{(-)}\rangle_{\ttb}$.  
As we push $w_c \to 1$, the cut-off slice folds into the bulk, eventually enveloping the bulk time-reflection symmetric slice in AdS. In this limit, the coefficient $Z^{(-)}_{UV}[w_c,g^{(0)}_-]$ can be viewed as the \emph{Hartle-Hawking wavefunction} of the CFT vacuum (up to possible counterterms)
\beq
\lim_{w_c \to 1} Z^{(-)}_{UV}[w_c,g^{(0)}_-] \sim \Psi^{(0)}[g^{(0)}_-]=\Psi^{(0)}[\gamma^{(\star)}],
\eeq
where $\gamma^{(\star)}$ is again the induced metric on the bulk time-reflection symmetric slice in AdS. 

So far, we have merely postulated that the states $|w_c,\gamma^{(-)}\rangle_{T\overline{T}}$ are CTN states, but we have not explained why this is a useful interpretation. As justification for our proposal, we will show in the next section that our proposed CTN states $|w_c, \gamma^{(-)}\rangle_{\ttb}$ satisfy a ``minimal-area'' bound on the entanglement entropies of all boundary intervals, with the coefficient being equal to $\frac{1}{4G_N}$. Furthermore, this bound gets saturated in the limit $w_c \to 1$, and so in this limit, the entanglement spectrum of these states becomes flat (at large $N$). 

\section{Entanglement entropy and the Ryu-Takayanagi formula}\label{sec:ee}
In this section, we discuss the entanglement structure of the CTN states in an attempt to justify their interpretation as tensor networks. 
\subsection{Bound on entanglement entropy}
\begin{figure}
    \centering
    \begin{tabular}{c c c}
    \includegraphics[height=3.2cm]{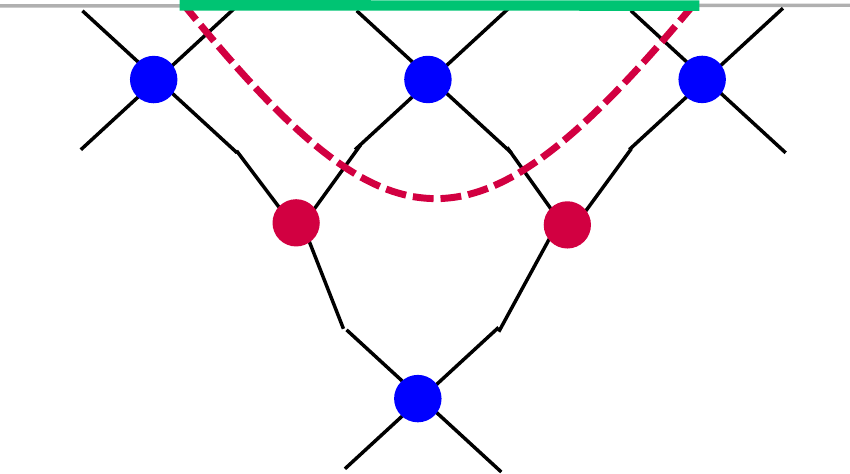} & \hspace{1cm} &\includegraphics[height=4.3cm]{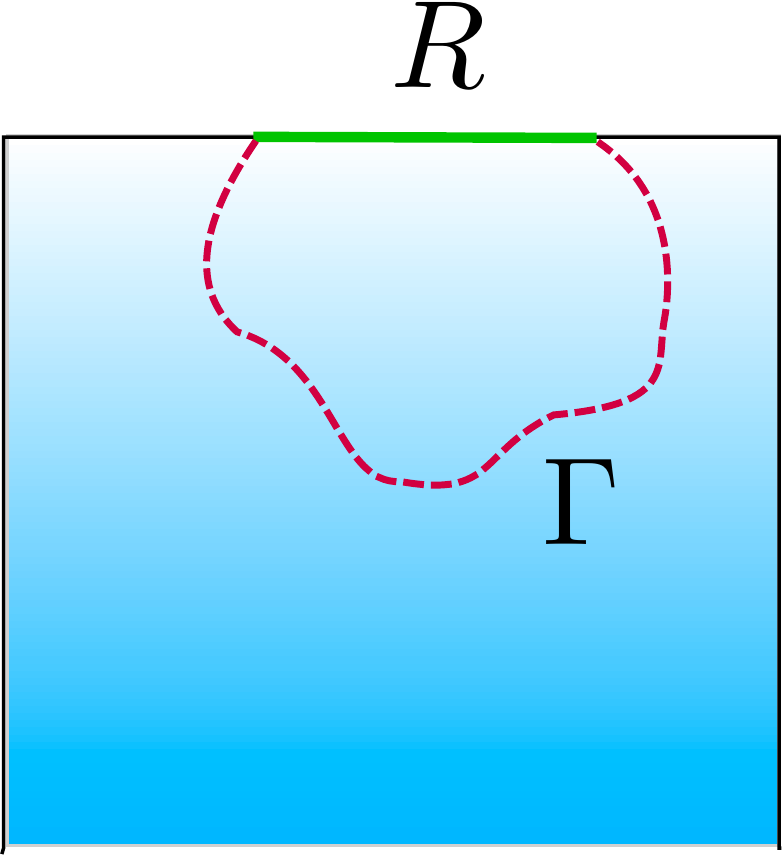}
    \end{tabular}
    \caption{\textbf{Left}: A cartoon for a tensor network preparation of a state. The entanglement entropy of some interval (shown in green) is upper bounded by the minimal cut (shown in red) through the network. \textbf{Right}: For the continuous case, the entanglement entropy of some subregion $R$ is upper bounded by $\log\,\text{dim}$ of the Hilbert space on any cut through the Euclidean path integral of the $\ttb$ deformed theory on the half place at $w_c$.}
    \label{fig:TN}
\end{figure}
An interesting feature of tensor networks is that they make manifest a natural, geometric, upper bound on the entanglement entropy of boundary intervals, which is rather reminiscent of the Ryu-Takayanagi formula. In the case of a discrete tensor network, the rough idea is as follows: we ask for the minimum number of cuts that one must make through the bonds in the tensor network, so as to completely dissociate the portion of the network containing the boundary interval in question from the rest of the network. Assuming that each bond has a \emph{bond dimension} (i.e., the dimension of the Hilbert space corresponding to the bond) $J$, then it follows that the entanglement entropy of the boundary interval must be bounded by
\beq
S_{EE} \leq (\log\,J) \;\times \text{min\;number\;of\;cuts}.
\eeq
If we interpret the minimal cut through the network as analogous to a ``minimal area surface'', then this bound is reminiscent of the Ryu-Takayanagi formula, the important difference being, of course, that the RT formula is an equality, not merely a bound. However, it has been shown that special tensor networks can be constructed which satisfy the RT formula as an equality \cite{Pastawski:2015qua, Hayden:2016cfa}. 

Our goal in this subsection is to argue that the CTN states $|w_c, \gamma^{(-)}\rangle_{\ttb}$ satisfy a similar bound on the entanglement entropy of boundary intervals. In our case, the network in question is not a discrete tensor network, but a continuous one, in the form of a Euclidean path integral. It is a general property of Euclidean path integrals in quantum field theory that we can cut them open along any co-dimension one slice of interest by inserting a complete set of states along that slice; we will assume that this is true in the $\ttb$ deformed theory. So, given some interval $R$ of length $L$ in the boundary CFT (at $\tau=0$ and $z=\epsilon$, which is also a boundary of the half-plane at $w_c$), we imagine cutting open the Euclidean path integral of the $\ttb$ deformed theory on the half plane at $w_c$ with metric $\gamma^{(-)}_{\mu\nu}$ along some curve $\Gamma$ which is homologous to $R$ and has the same end points (see figure \ref{fig:TN}). For any such $\Gamma$, it must be that the entanglement entropy of $R$ is upper bounded by
\beq
S_{EE}(R) \leq \log\, \text{dim}\, \mathcal{H}_{\Gamma},
\eeq
where $\mathcal{H}_{\Gamma}$ is the Hilbert space on $\Gamma$. Of course, the best bound is obtained by picking a curve such that the right hand side is minimized, so that
\beq \label{bound1}
S_{EE}(R) \leq \text{min}_{\Gamma}\,\log\, \text{dim}\, \mathcal{H}_{\Gamma}.
\eeq
Now in most standard quantum field theories, this upper bound is not very useful, because the dimension of the Hilbert space tends to be infinite. One way to see this, for example, is to compute the thermal partition function in the limit $\beta \to 0$, which computes the dimension of the field theory Hilbert space. In a 2d CFT on a circle for instance, this object diverges as 
\beq
\dim \mathcal{H}_{\Gamma} = \lim_{\beta \to 0} Z_{CFT}(T^2) = \lim_{\beta \to 0} \mathrm{Tr}\,e^{-\beta\, H} \sim e^{\frac{c}{12}\frac{2\pi L_{\Gamma}}{\beta}},
\eeq
where $L_{\Gamma}$ is the length of the spatial circle. The same is also true of the CFT on an interval. However, we seem to be in a better situation, as the path-integral of interest for CTN states is that of a $T\overline{T}$ deformed quantum field theory, and the $\ttb$ deformation acts as a natural regulator for this divergence. Thus, in the present case, we can hope to achieve a meaningful upper bound. 

In order to compute the dimension of the Hilbert space of the deformed theory on $\Gamma$, we begin by noting that in a tubular neighborhood of $\Gamma$, we can always pick an adapted coordinate system $(\sigma_0,\sigma_1)$ in which the metric looks flat, up to corrections from the extrinsic curvature $K$ of $\Gamma$. Here, $\sigma_1 \in [0, L_{\Gamma}]$ is the proper length along $\Gamma$ and $\sigma_0$ is a Euclidean time coordinate, such that the metric comes:
$$ ds^2 = d\sigma_0^2 + d\sigma_1^2 + O(K\sigma^0).$$
Assuming differomorphism covariance, the action/measure for the Euclidean path-integral in this tubular neighbourhood should therefore look locally like that of the flat-space $\ttb$-deformed theory with coupling $\lambda = 2\pi G_N w_c^2$, up to sub-leading corrections from the extrinsic curvature; all the remaining metric dependence is now packaged into the length $L_{\Gamma}$, which is of course computed using the metric $\gamma^{(-)}$.  We can therefore obtain the dimension of the Hilbert space on $\Gamma$ by using this description, i.e.,  by computing the cylinder partition function $Z_{\ttb}(\beta, L_{\G}, \l)$ in the flat-space $\ttb$-deformed CFT  for fixed spatial length $L_{\Gamma}$ and $\ttb$ coupling $\lambda$, with the inverse temperature $\beta \to 0$. 

We can do this computation in two ways: 1. using a field theory argument involving the spectral flow of the ground state energy,\footnote{We thank John Cardy for a helpful exchange on this method.} or 2. using AdS/CFT. Here we will present the field theory method, leaving the AdS/CFT computation to Appendix \ref{sec:dimH}. Using a global scale transformation, we can write
\beq
Z_{\ttb}(\beta, L_{\G}, \l) = Z_{\ttb}\left(1, \frac{L_{\G}}{\beta},\frac{\l}{\beta^2}\right).
\eeq
In the torus case, we could alternatively use the modular invariance properties of the $\ttb$-deformed torus partition function \cite{Datta:2018thy, Aharony:2018bad}. In the limit $\beta \to 0$, the effective length of the cylinder in the partition function on the RHS is going to infinity. It is therefore convenient to quantize by treating the periodic direction as space and the non-compact direction as time. As $\beta \to 0$ we can thus write
\beq\label{TTPF0}
\lim_{\beta \to 0} Z_{\ttb}(\beta, L_{\G}, \l) \sim \exp\left[-\frac{L_{\G}}{\beta}E^{(0)}\left(1,\frac{\l}{\beta^2}\right)\right],
\eeq
where $E^{(0)}(1,\frac{\l}{\beta^2})$ is the vacuum energy of the $\ttb$-deformed theory on a spatial circle of length $1$ and coupling $\l/\beta^2$ \footnote{The viewpoint that we are taking here is that states with complex energy should be excluded from the spectrum. This means that once an energy level goes complex, we project it out. One subtlety here is that for non-zero angular momentum, there are energies that go below the deformed vacuum energy, but those are not real for all $\l$ and would hence be excluded from the spectrum. In the bulk they correspond to a situation with two real horizons.}. Using the Burgers' equation solution for the deformed energy eigenvalues equation \eqref{EFlow}, we have 
\beq
E^{(0)}\left(1,\frac{\l}{\beta^2}\right)= \frac{\beta^2}{4\l}\left(1-\sqrt{1-\frac{8\l E^{(0)}(1,0)}{\beta^2}}\right),
\eeq
where $E^{(0)}(1,0) = -\frac{\pi c}{6}$ is the vacuum energy in the original undeformed CFT. In the $\beta \to 0$ limit, this gives
\beq
\lim_{\beta \to 0}E^{(0)}\left(1,\frac{\l}{\beta^2}\right)= -\beta\sqrt{\frac{\pi c} {12\l}} + O(\beta^2).
\eeq
Substituting this into equation \eqref{TTPF0}, we get
\beq\label{TTPF}
\dim \mathcal{H}_{\Gamma} = \lim_{\beta \to 0}Z_{\ttb}(\beta, L_{\G}, \l) \sim \exp\left(\sqrt{\frac{\pi c} {12\l}}L_{\G}\right).
\eeq
Since the effective length of the cylinder is going to infinity, we also expect that the details of boundary conditions on the cylinder are not important, at least at large $c$. In addition, as noted before we could also do this computation using AdS/CFT and the MMV conjecture \cite{McGough:2016lol}; this has been done in Appendix \ref{sec:dimH} and agrees precisely with the above result.  

Returning to our calculation of the dimension of the Hilbert space of a $T\overline{T}$ deformed theory, we find, taking the log of \eqref{TTPF} and using $\l = 2\pi G_N w_c^2$ and $c = 3/2G_N$,
\beq \label{bound2}
\log\, \text{dim}\, \mathcal{H}_{\Gamma}= \frac{1}{4G_N w_c}L_{\Gamma}(\gamma^{(-)}) =\frac{1}{4G_N }L_{\Gamma}(g^{(0)}_-) ,
\eeq
where $L_{\Gamma}(g^{(0)}_-)$ is the length of the curve $\Gamma$ computed using the induced metric $g^{(0)}$. This formula suggests that the $\ttb$ deformed field theory regulates the local Hilbert space dimension (i.e., entropy density) with respect to $g^{(0)}_-$ to be finite and equal to $\frac{1}{4G_N}$.\footnote{A similar observation was also made in \cite{Donnelly:2018bef} using the $n\to 0$ limit of the Renyi entropy.} Now, the crucial point is that using equations \eqref{bound1} and \eqref{bound2}, we get a finite upper bound on the entropy of boundary intervals
\beq
S_{EE}(R) \leq \frac{1}{4G_N} \text{min}_{\Gamma}\,L_{\Gamma}(g^{(0)}_-),
\eeq
In the limit $w_c \to 1$, we then get
\beq
S_{EE}(R) \leq \frac{1}{4G_N} \text{min}_{\Gamma}\,L_{\Gamma}(\gamma^{(\star)}),
\eeq
where as before $\gamma^{(\star)}$ is the induced metric on the bulk time-reflection symmetric slice. Thus we recover the Ryu-Takayanagi-like bound on the entropy. In fact, the coefficient works out to be precisely $\frac{1}{4G_N}$. We may interpret this coefficient as the log of the effective bond dimension, if we assume that individual tensors have a small but $O(1)$ size. On the other hand, since the natural length scale of non-locality is $\sqrt{\l}$, it seems more natural to interpret the tensors as having $O(\sqrt{\l})$ thickness, with the individual bond dimension being $O(\sqrt{c})$. This fine-grained structure of the tensors is consistent with the expectations \cite{Swingle:2012wq, Bao:2015uaa} of sub-AdS locality \cite{Heemskerk:2009pn} in networks dual to states in a strongly coupled holographic theory, and also fits well the with complexity=volume conjecture \cite{Susskind:2014rva, Stanford:2014jda}, as we will discuss further in the discussion section. 

We should emphasize that although the finite upper bound in the present case falls out of the computation of the torus/cylinder partition function in the $\ttb$ deformed field theory, the underlying reason for this finiteness is, to our knowledge, not completely understood. It has recently been argued that the $\ttb$ deformation gives elementary excitations in the theory a finite width \cite{Cardy:2020olv, Jiang:2020nnb}. This may ultimately be the physical reason for the finite entropy density. The other point to emphasize is that the computations we have relied on take the $N\to \infty$ limit \emph{first} and then $\beta \to 0$; in other words, the Hilbert space dimension may yet have infinities which are subleading in $1/N$.\footnote{Of course, the length of the minimal length curve with respect to $\gamma^{(\star)}$ also diverges as we send $\epsilon \to 0$.}

\subsection{Computing the entropy at large $N$}

In the previous section, we obtained an upper bound on the entanglement entropy of boundary intervals. Here we wish to do better and actually compute the entanglement entropy in the CTN states $|w_c,\gamma^{(-)}\rangle$ at large $N$. We will only do this at large $N$ in the saddle point limit, where the state is prepared by the path integral of the deformed theory on the background metric $\gamma^{(-)}(w_c)$ given by:
\beq
\gamma^{(-)}(w_c) = \frac{f(w_c)}{w_c^2}\frac{(du^2+dx^2)}{u^2},\;\;\; |u| > u_0,
\eeq
with a thin, regulator strip of flat space of width $b=\frac{1-w_c^2}{2w_c}\epsilon$ in the time direction attached at $u=u_0$. In the limit $w_c \to 1$, this metric approaches that of the bulk time reflection symmetric slice, which as before will be referred to as $\gamma^{(\star)}$:
\beq
\gamma^{(\star)} = \frac{(du^2+dx^2)}{u^2},\;\;\; |u| > \epsilon,
\eeq
with the width of the regulator strip going to zero in this limit. We are interested in computing the entanglement entropy of some boundary interval $R$ of length $L$. The simplest way to get the answer is to use the holographic Ryu-Takayanagi formula. Since we are in a time-reflection symmetric setup, the Ryu-Takayanagi surface for an interval at $\tau=0$ (and $z=\epsilon$) in the boundary must lie along the time-reflection symmetric slice in the bulk. Further, note that the $\ttb$ deformation in question has no effect on the geometry of the bulk time-reflection symmetric slice. Therefore, for any value of $w_c$, the Ryu-Takayanagi surface and the corresponding entanglement entropy remain unchanged:
\beq
S_{EE}(R) = \frac{1}{4G_N} \text{min}_{\Gamma}L_{\Gamma}(\gamma_{\star}).
\eeq
\begin{figure}
    \centering
    \includegraphics[height=5.5cm]{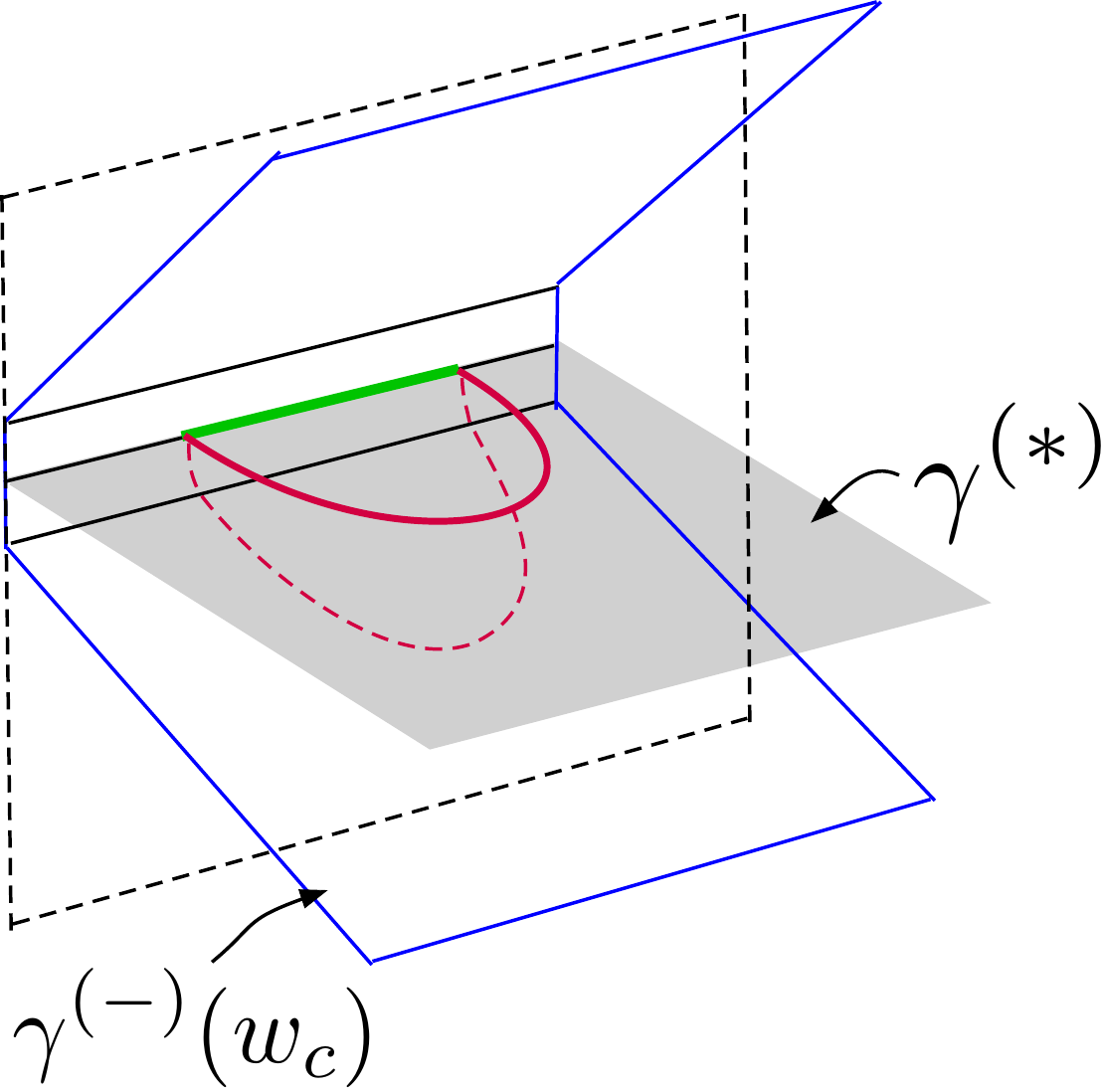}
    \hspace{1cm}
    \includegraphics[height=5cm]{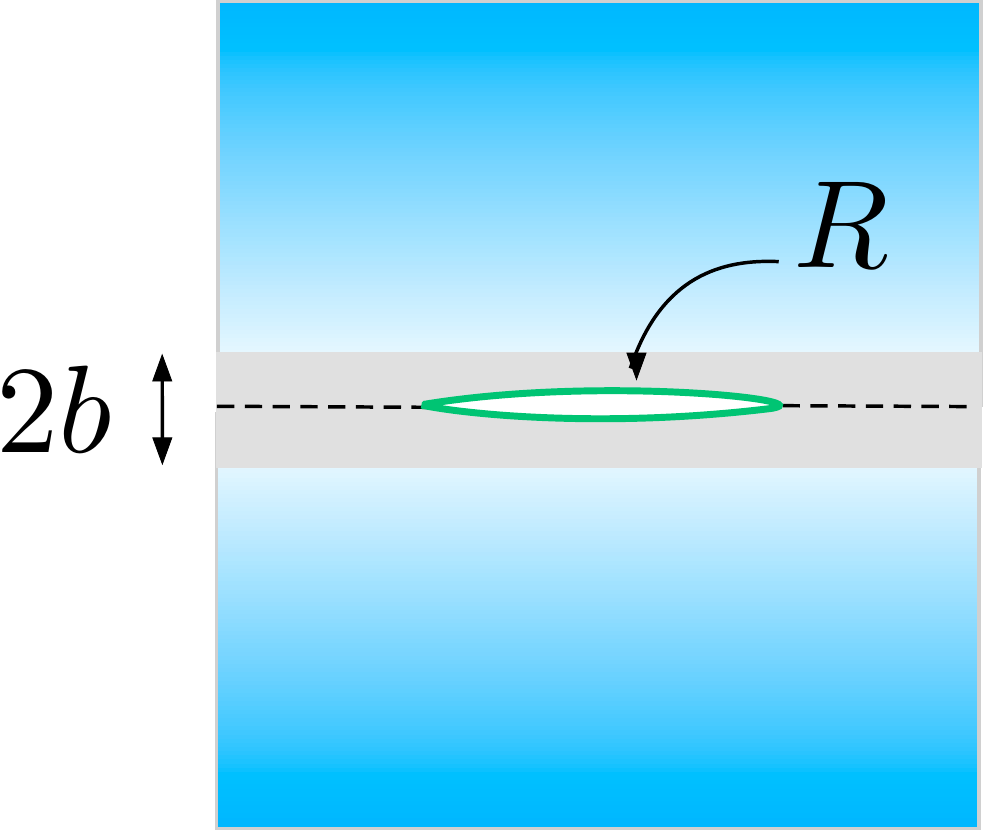}
    \caption{\textbf{Left}: Folded slices $\g^{(-)}(w_c)$ and $\g^{(*)}$. $\g^{(*)}$ is the metric on the time-reflection symmetric slice in the bulk, which is shaded gray. The blue slice is a particular folded slice at some $w_c$. The green region is the boundary region $R$ of which we want to compute the entanglement entropy. The dashed red curve is the minimal-rank cut on the constant $w_c$ slice and the solid red curve is the actual RT surface on the bulk time-reflection symmetric slices $(w_c = 1)$. \textbf{Right}: The path-integral computing the density matrix on $R$. The regulator strip of size $2b$ is shown in grey. The replica manifold $\mathcal{M}_n$ is obtained by taking the $n$-fold branched cover over $R$. }
    \label{fig:RT}
\end{figure}
Note that this is only true of the leading, large $N$ contribution to the entropy. The subleading contributions will in general depend on $w_c$, but will not be considered here. So, the picture we have for the entropy is as follows:  as we flow along the deformation parameter $w_c$, the large $N$ entropy of a boundary interval stays constant; i.e., the one parameter family of tensor network states labelled by $w_c$ all have the same entropies at large $N$. However, the minimal-cut upper bound on the entropy we derived in the previous section becomes smaller and smaller (i.e., more and more constraining) as we dial up the $\ttb$ coupling, and eventually as we approach $w_c \to 1$, the bound saturates to the Ryu-Takayanagi formula for the entanglement entropy (see left panel of figure \ref{fig:RT}). Since that bound came from the rank of the density matrix, we conclude that in this limit the entropy is maximal. Therefore, we expect the entanglement spectrum to be \emph{flat}, i.e., all the eigenvalues of the density matrix are equal to $1/\dim \mathcal{H}$, up to $1/N$ corrections. This is indeed a standard property of many of the tensor network models which have been constructed in the literature. Recently, a class of states called \emph{fixed-area} states were introduced in \cite{Dong:2018seb, Akers:2018fow} to capture this property of tensor networks. The construction given in these papers used the bulk gravity description, but a field theory description of fixed-area states is not known, to the best of our knowledge. Our CTN states in the limit $w_c \to 1 $ are then a natural candidate for a field theory description of fixed-area states. 

We can also give a field theory argument for the entanglement entropy being independent of $w_c$, using the trace flow equation. Following \cite{Donnelly:2018bef, Lewkowycz:2019xse}, we consider the R\'enyi entropy:
\beq
S_n = -\frac{1}{n-1}\,\log\,\mathrm{Tr}_R \rho_R^n.
\eeq
The entanglement entropy is obtained from the R\'enyi entropy by analytically continuing in $n$ and taking the limit $n\to 1$. The R\'enyi entropy for integer $n$ can be computed by using the replica trick:
\beq
\mathrm{Tr}_R \rho_R^n = \frac{Z_n}{Z_1^n},
\eeq
where $Z_n$ is the Euclidean path integral on the replica manifold $\mathcal{M}_n$ (see the right panel of figure \ref{fig:RT}). 

The idea is to now consider the derivative of the R\'enyi entropy with respect to the length of the interval:
\beq
L\frac{d}{dL}S_n = -\frac{1}{n-1}\left[\int_{\mathcal{M}_n} d^2x\,\langle {T^{\mu}}_{\mu}\rangle_{\mathcal{M}_n} - n\int_{\mathcal{M}_1} d^2x \,\langle{T^{\mu}}_{\mu}\rangle_{\mathcal{M}_1}\right].
\eeq
Using the $\mathbb{Z}_n$ symmetry of the replica manifold, we can write the first term above -- which at present is an integral over all the $n$ sheets -- as an integral over one of the sheets:
\beqn
\int_{\mathcal{M}_n} d^2x\,\langle {T^{\mu}}_{\mu}\rangle_{\mathcal{M}_n}&=& \sum_{i=1}^n\int_{\mathcal{M}_1} d^2x_i\,\langle {T^{\mu}}_{\mu}(x_i)\rangle_{\mathcal{M}_n}\nonumber\\
&=& n\int_{\mathcal{M}_1} d^2x\,\langle {T^{\mu}}_{\mu}(x)\rangle_{\mathcal{M}_n}.
\eeqn
Therefore, we get
\beq
L\frac{d}{dL}S_n = -\frac{n}{n-1}\int_{\mathcal{M}_1} d^2x\,\Big(\langle {T^{\mu}}_{\mu}\rangle_{\mathcal{M}_n} - \langle{T^{\mu}}_{\mu}\rangle_{\mathcal{M}_1}\Big).
\eeq
In \cite{Donnelly:2018bef, Lewkowycz:2019xse}, the authors consider situations which have a rotation symmetry around the entanglement cut, together with the trace flow equation and conservation equations to solve for the above one point $\langle {T^{\mu}}_{\mu}\rangle_{\mathcal{M}_n}$ in the large $N$ limit. In the present case, the density matrix corresponding to the CTN state $|w_c, \gamma^{(-)}(w_c)\rangle$ does not have this rotation symmetry, and thus the solution for the stress tensor is not simple to obtain. However, we are only interested here in the entanglement entropy, for which we can take the $n\to 1$ limit. In this limit, we only need to compute $\langle {T^{\mu}}_{\mu}\rangle_{\mathcal{M}_n}$ to linear order in $\alpha = (n-1)$. This, we can do at large $N$, by appealing to the trace flow equation:
\beq
{T^{\mu}}_{\mu}=-4\pi G_N\Big( T_{\mu\nu} T^{\mu\nu}- ({T^{\mu}}_{\mu})^2\Big) -\frac{1}{16\pi G_N} R_{\gamma_{\alpha}},
\eeq
where all the stress tensors above are understood to be one-point functions on $\mathcal{M}_n$. Now consider a point $x$ on $\mathcal{M}_n$ which is far from the entanglement cut (i.e., the branch points of the replica manifold). Taking a variation of this equation with respect to $\alpha$, we get
\beq \label{TrDer}
\delta_{\alpha}{T^{\mu}}_{\mu}=-8\pi G_N\Big( T^{\mu\nu} \delta_{\alpha} T_{\mu\nu}- {T^{\mu}}_{\mu}\delta_{\alpha}{T^{\mu}}_{\mu}\Big).
\eeq
where we have used the fact that the metric on the replica manifold is locally $\alpha$-independent for $x$ far from the entanglement cut. For the same reason, we have also dropped the $\delta_{\alpha} R$ term above. Now we set $\alpha=0$ (i.e., $n=1$) in the above equation. In order to compute the local stress tensor on $\mathcal{M}_1$, we can use the fact that the geometry is homogeneous away from the kink at $|u|=u_0$; thus away from this kink, we can make the ansatz:
\beq \label{BgT}
T^{(\alpha=0)}_{\mu\nu} = \varepsilon\,\gamma_{\mu\nu},
\eeq
where $\varepsilon$ is locally constant. Using the trace flow equation on $\mathcal{M}_1$, we get
\beq
\varepsilon = \frac{1}{8\pi G_N w_c^2}\left(1- \sqrt{1-\frac{w_c^2}{r^2(w_c)}}\right)
\eeq
in the hyperbolic region $|u| > u_0$, where $r(w_c)$ is the radius of curvature of this hyperbolic region: $R_{\gamma} = -\frac{2}{r^2}$. In the flat regulator strip, we find $\varepsilon=0$. Of course, this solution is only valid away from the kink at $|u|=u_0$; in a neighborhood of the kink we can smooth out the geometry by hand and then solve for the local stress tensor. Since the kink is a measure zero set, we do not expect the stress tensor to be singular in this region, nevertheless it would be interesting to check this expectation in detail. At any rate, having found $T^{(\alpha=0)}$ (at least away from the kink), we can now go back to equation \eqref{TrDer}, and using equation \eqref{BgT} we thus conclude that 
\beq
\delta_{\alpha}{T^{\mu}}_{\mu}(x)=0,
\eeq
for any point $x$ far from the entanglement cut (and the kink). Thus the entire contribution to the entanglement entropy must come from an infinitesimal neighborhood around the entanglement cut -- this is very similar to what happens in the rotation-symmetric case of \cite{Donnelly:2018bef, Lewkowycz:2019xse}. Crucially, the $\ttb$ deformation we are presently considering turns off in a strip of width $2b$ around the entanglement cut. Therefore, the entanglement entropy will not change as we deform along the parameter $w_c$. This is indeed consistent with what the Ryu-Takayanagi formula also predicts, as discussed above. While we have not been able to explicitly compute the R\'enyi entropies using the trace flow equation, the saturation of the rank bound in the $w_c\to 1$ limit suggests that at least in this limit, all the R\'enyi entropies are equal. It will be interesting to confirm this with an explicit calculation. 

\section{Discussion and further developments}
 In this section, we will discuss some potential applications of our construction and future directions, and then end with a brief discussion of the loose ends.

 \subsection{Sub-AdS locality and Complexity $=$ Volume}
Since we have a continuous tensor network interpretation of bulk slices, we could ask whether this sheds any light on the complexity = volume conjecture \cite{Susskind:2014rva, Stanford:2014jda}, which states that the volume of the maximal volume slice anchored at boundary time $t$ in Lorentzian signature computes the complexity of the boundary CFT state at $t$. Of course, we do not yet have a way to extend our network interpretation to Lorentzian signature, but we could nevertheless consider the slice at $t=0$, i.e., the time-reflection symmetric slice in Euclidean signature. One obvious problem is that our tensor network is a continuous one, in the form of a Euclidean path integral. However, the path integral is performed in the $\ttb$-deformed field theory, which has a natural length scale of non-locality $\ell_{NL} \sim \sqrt{\l}$ (see, for example, \cite{Cardy:2020olv, Jiang:2020nnb}). With this in mind, it seems natural to think of an elementary tensor as occupying a region of size $\ell_{NL}^2$ in the path integral. If we let the complexity of the network at $w_c$ -- which we denote $\mathcal{C}(w_c)$ -- mean the total number of tensors in the network, then we get (see also \cite{Geng:2019yxo} for a similar argument):
\beq
\mathcal{C}(w_c) = \int d^2x\; c(w_c,x),\;\;\; c(w_c,x) = \frac{\sqrt{\text{det}\,\gamma_{\mu\nu}^{(-)}(w_c,x)}}{\ell_{NL}^2}.
\eeq
Taking $\ell_{NL} \sim \sqrt{\l} = \sqrt{2\pi G_N w_c^2}$ and writing the above expression in terms of the induced metric $g^{(0)} = \frac{1}{w_c^2}\gamma^{(-)}$, we find
\beq
\mathcal{C}(w_c) \sim \frac{1}{2\pi G_N}\int d^2x\, \sqrt{\text{det}\,g_{\mu\nu}^{(0)}(w_c,x)} = \frac{1}{2\pi G_N} \text{Volume}(w_c).
\eeq
Indeed, this volume is minimized on the time-reflection symmetric slice $w_c=1$, and hence we conclude that the corresponding network is the optimal tensor network within the family we have considered in this work. It has indeed been suggested previously \cite{Caputa:2017yrh, Belin:2018bpg} that the complexity = volume conjecture should be interpreted in the context of Euclidean path integrals (as opposed to unitary circuit complexity), and our considerations here support this idea. Note that it was crucial that the length scale of non-locality in the network is proportional to $\sqrt{G_N}$ (and not $\ell_{AdS}$) in order to get the factor of $\frac{1}{G_N}$ in the complexity.

\subsection{Bit threads}
Bit threads were proposed by Freedman and Headrick \cite{Freedman:2016zud} (see also \cite{Headrick:2017ucz}) as an alternate formulation of the holographic entanglement entropy formula of Ryu and Takayanagi. Their proposal is rooted in a concept in network theory, called min-cut-max-flow, and formulates the search for a minimal surface (the RT surface) in terms of the maximation of a flow on a Cauchy slice in the bulk. Specifically, consider a vector field $v^\mu$ on this Cauchy slice $\Sigma$ in the bulk that has the following two properties
\be 
\nabla_\mu v^\mu = 0, \quad |v| \leq C,
\ee
for a positive constant $C$. Let us consider a region $R$ at the boundary of $\Sigma$. The min-cut-max-flow theorem is then the statement that
\be 
\underset{v}{\rm max} \int_R v = C\;\underset{m\sim R}{{\rm min}}\;A(m),
\ee
with $m\sim R$ a codimension-one surface $m$ homologous to $R$ and $A(m)$ its area. The RHS is the well-known RT formula when we take $\Sigma$ to be the time-reflection symmetric slice and we set $C = 1/4G_N$. In this case, the field lines of the vector field $v$ (bit threads) have a ``finite size'' in Planck units, so to speak. 

It is interesting to ask whether we can give a physical interpretation for bit-threads in the context of our tensor network. One possible interpretation is that the Euclidean path-integral of the $\ttb$ deformed theory in the limit $w_c\to 1$ should be thought of as a ``Euclidean fluid''.  
  It is then natural to ask whether we can define an \emph{entropy current}, $J_S^{\mu} = s\, u^{\mu}$, where $s$ is the local entropy density, and we have introduced an arbitrary vector field $u^{\mu}$ with norm $|u|=1$, which we think of as the fluid velocity profile. For a given boundary interval $R$, we would like to take $u^{\mu}$ to be a flow from $R$ to $\bar{R}$. It is clear that the local entropy density of this fluid should be upper bounded by the rank of the local Hilbert space, which as shown in section \ref{sec:ee} and Appendix \ref{sec:dimH}, is given by $\frac{1}{4G_N}$. This then implies
\beq
|J_S|  = s \leq \frac{1}{4G_N}.
\eeq
Furthermore, in usual fluid dynamics, the entropy current satisfies $\nabla_{\mu}J_S^{\mu} \geq 0$. However, if our fluid is not dissipative, then we would expect that
\beq
\nabla_{\mu}J_S^{\mu} =0.
\eeq
These are indeed the conditions satisfied by bit-thread flows $v^{\mu}.$ Of course, the flux of $J_S^{\mu}$ through the minimal area surface (or any other homologous surface) is the total coarse-grained entropy of the fluid configuration, and so the min-cut-max-flow theorem says that the entanglement entropy is the maximum coarse-grained entropy (over all possible choices of $u^{\mu}$) that this $\ttb$ fluid flowing between $R$ and its complement can have. It would be interesting to further explore this interpretation. 


\subsection{Tensor networks in real time}
There has been some debate in the context of tensor networks in AdS/CFT on whether the network should be placed on either a space-slice, time-like or null slice in the bulk and accordingly have a Euclidean, Lorentzian or degenerate intrinsic geometry \cite{Czech:2015kbp, Milsted:2018san}. Our proposal suggests that the (space-like) time-reflection symmetric slice in the bulk can be interpreted a Euclidean tensor network. An interesting question is whether we can extend our proposal to surfaces in Lorentzian geometry. For instance, in Lorentizan AdS spacetime, we could consider a dS slicing of the bulk geometry. In this case, we may interpret the Lorentzian path integral of the $\ttb$-deformed boundary field theory as preparing a tensor network state via \emph{unitary} (i.e., real-time) evolution on a de-Sitter geometry, starting from some fiducial state at large negative time. 


Another natural question is whether we can take our Euclidean network on the time-reflection symmetric slice, and extend it to Lorentzian signature by flowing in real time using the Wheeler-de Witt equation. It seems reasonable to think that this also corresponds to a modified $\ttb$ flow in the effective theory on the time slices. More explicitly, the Wheeler-de Witt equation in Lorentzian signature with real time treated as the flow direction is given by (compare to \eqref{radialH})
\be 
K^2 - K_{ab}K^{ab} + (R_{g^{(0)}} + 2) = 0,
\ee
with $g^{(0)}$ the induced metric on the space-like slices. Since the Balasubramanian-Kraus stress-tensor is defined at each slice as in \eqref{TBK}, and assuming that the counter term piece is unaltered, we arrive at the following flow equation for the bulk Balasubramanian-Kraus stress tensor,
\be\label{TFET}
T_a^a = - 4\pi G_N (T_{ab}T^{ab} - (T_a^a)^2) + \frac{1}{4\pi G_N} + \frac{R_{g^{(0)}}}{16\pi G_N},
\ee
where the index contractions are with respect to $g^{(0)}$. It is important to notice that at the bulk time-reflection symmetric slice this trace flow equation conincides with \eqref{TFE}, since this slice has $R_{g^{(0)}} = -2$ so that the last two terms combine into $1/8\pi G_N$, in agreement with the last term in \eqref{TFE} which has $R_{\g} = -2$. Therefore, we could first flow with the ``folding deformation'' in Euclidean signature till the time-reflection symmetric slice, and then continue into Lorentzian signature with the flow described above; this gives a tensor network interpretation to slices in Lorentzian signature. The third term on the RHS in equation \eqref{TFET} is similar to the $\Lambda_2$ deformation proposed in \cite{Gorbenko:2018oov}. The fourth term may seem odd, because it has the opposite sign, but this is required for a consistent gluing of the two flows. In this two stage flow, we can think of the Euclidean flow as generating the initial conditions for the subsequent flow in real time. It would be interesting to explore this in more detail, and in particular explore connections with \cite{ May:2016dgv, Mezei:2018jco}. This line of thought may also have interesting applications in cosmological settings and dS/CFT. We hope to return to this in the near future.  
    
\subsection{Other generalizations and loose ends}
There are various further generalizations we could imagine. So far we have applied our tensor network construction to a two-dimensional conformal field theory, but we could equally well apply it to other dimensions. In higher dimensions, the argument of the bound on the entanglement entropy for a region becomes much more intricate. On the other hand, it may be simpler to consider the lower-dimensional case of JT gravity in two bulk dimensions, where we expect to have the same construction. Specifically, the folded slices in the bulk are dual to turning on the deformation of \cite{Gross:2019ach, Gross:2019uxi} in the putative boundary quantum mechanics. One could for instance consider the low-temperature SYK path integral over some Euclidean time $\t \leq 0$ \footnote{There is no time-dependence in the one-dimensional metric as we can simply redefine our time coordinate}. Adding the deformation of \cite{Gross:2019ach, Gross:2019uxi} would then be dual to putting the $1d$ theory on the folded slices, at least semi-classically. Following the deformation all the way to the bulk time-reflection symmetric slice then gives a tensor network state representation of the ground state of SYK, just as we had in equation \eqref{tn1}. Another interesting situation to consider is the thermofield double state, where it would be interesting to understand the entropy (computed in JT gravity in terms of the dilaton) from our tensor network perspective. It seems clear that our arguments should also go through in the case of the thermofield double state in three bulk dimensions, at least in the black hole phase. 

We have also restricted our considerations here to large $N$. A natural next step would be to include the leading $1/N$ corrections, which are related to quantum entanglement in the bulk \cite{Faulkner:2013ana}. Relatedly, our discussion has focused on tensor network \emph{states} and it would be interesting to generalize this to holographic quantum codes, which also include bulk degrees of freedom \cite{Pastawski:2015qua}. Finally, it is also worth exploring entanglement wedge reconstruction from our perspective.

While our construction of the CTN states provides a framework to formalize various aspects of the AdS/tensor network correspondence, we should note here that the subject of $\ttb$ deformations of quantum field theories is not completely settled. In particular, the $\ttb$ operator is irrelevant, and at least with the holographic sign, naively leads to a complexification of the energy spectrum beyond the energy $\frac{1}{4\lambda}$ (at zero angular momentum), where $\lambda$ is the $\ttb$ coupling. This is an indication that the deformation drastically modifies the UV limit of the theory and good UV completions are not known. Thus, in order to really make the AdS/tensor network correspondence precise along the lines we have suggested, we need to have good UV completions for $\ttb$ deformed holographic CFTs. In addition, we also need to understand how the $\ttb$ deformation works on curved spaces at finite $N$.  

Having said that, we will end on the following optimistic note: there is something essentially discrete about the Ryu-Takayanagi formula and its tensor network incarnations which have appeared in the literature. Indeed, the RT formula suggests a pixelated picture of the ``fabric of spacetime'', with $\frac{1}{4G_N}$ being the information content of an individual pixel on the RT surface. Various tensor network models have captured this essential discreteness in a nice way. But given a holographic CFT, there has so far been no principled way to construct the discrete elements/tensors out of which to build the tensor network.  Our proposal here supplies a partial answer -- the $\ttb$ deformation flows towards a potential UV theory which completes the deformed CFT. The tensors which make up the ``fabric of spacetime'' are then pieces of the Euclidean path integral in this UV theory. 

\section*{Acknowledgements}
We thank Alex Belin, Jan Boruch, John Cardy, Dongsheng Ge, Paolo Glorioso, Yuri Lensky, Mark Mezei, Xiaoliang Qi, Gábor Sárosi, Jonathan Sorce and Tadashi Takayanagi for helpful discussions and comments on an earlier version of the manuscript. JK is supported by the Simons Foundation. PC is supported by NAWA “Polish Returns 2019” and NCN Sonata Bis 9 grants.
\appendix

\section{Details of the folding deformation}\label{Details}
In this appendix, we give some further technical details about the folding $\ttb$ deformation which was discussed in the main text. 

We start with the AdS$_3$ spacetime in Poincaré coordinates 
\be
ds^2=\frac{dz^2+d\tau^2+dx^2}{z^2},
\ee
and will be interested in the computation of the holographic stress tensor as well as the on-shell action on the wedge geometry. For this, it will be convenient to go to polar coordinates in the $(\tau,z)$ plane:
\be
z=u \cos\theta,\qquad \tau=u\sin\theta,\qquad \theta\in \left[-\frac{\pi}{2},\frac{\pi}{2}\right],\qquad u\in [0,\infty).
\ee
The coordinate $\theta$ is related to the coordinate $w$ which was used in the main text by
\beq
\cos\theta = \frac{2w}{1+w^2}.\label{wteta}
\eeq
We would like to consider $w$ (or $\theta$) as the ``radial'' coordinate, and consider the boundary field theory as living at finite $w$. Note that for a given $w$, we can solve for $\theta(w)$ using the above relation, but there are two solutions, i.e., $\theta(w)$ and $-\theta(w)$. These correspond to the upper half plane and the lower half plane of the boundary field theory, and we can equivalently describe the them by positive or negative $u$, respectively, if we wish to do so. At any rate, in these coordinates the metric becomes
\bea
ds^2= \frac{d\theta^2}{\cos^2\theta}+\frac{1}{\cos^2\theta}\frac{du^2+dx^2}{u^2}.
\eea
In what follows, we will be interested in an effective theory described by the region of the bulk geometry confined to the wedge between $-\theta_c\le\theta\le\theta_c$. See similar computations in the context of recently-proposed wedge holography \cite{Akal:2020wfl}. Since we will be interested in the region of space-time up to $z=\epsilon$, coordinate $u$ will be restricted to
\be
u\in \left[u_0\equiv\frac{\epsilon}{\cos\theta},\infty \right).
\ee

We start with computing expectation values of the holographic stress tensors and their trace. For $\theta=\theta_c$ we have the (outward pointing) normal vectors
\be
n_\mu=s\frac{1}{\cos\theta}\delta_{\mu\theta},\qquad n^\mu=s \cos\theta \delta^{\mu \theta},\qquad n_\mu n^\mu=1.
\ee
where $s=+1$ for positive $\theta_c$ and $s=-1$ negative for negative $\theta_c$ boundaries. The extrinsic curvature and its trace are given by
\be
K_{ij}=e^\mu_i e^\nu_j \nabla_\mu n_\nu=s \sin\theta h_{ij}=\frac{1}{2}Kh_{ij},\qquad K=h^{ij}K_{ij}=2s\sin\theta,
\ee
where $h_{ij}$ is the induced metric at constant $\theta = \theta_c$ slices. Clearly, these slices are the ``constant meant curvature" (CMC) slices of AdS$_3$ and they also have constant negative Ricci scalar 
\be
R_h = -2\cos^2\theta_c .
\ee
With this data, we can also confirm that the Hamiltonian constraint of GR on our slices is satisfied
\be
K^2-K^{ij}K_{ij}=2\sin^2\theta_c = R_h-2\Lambda.
\ee
The holographic stress tensors on the boundaries of the wedge are given by
\be
T_{ij}=\frac{1}{\kappa^2}\left(K_{ij}-Kh_{ij}+h_{ij}\right)=\frac{1}{2\kappa^2}\left(2-K\right)h_{ij},
\ee
and their trace is
\be
T^i_i=h^{ij}
T_{ij}=\frac{1}{\kappa^2}\left(2-K\right)=\frac{2}{\kappa^2}\left(1-\sin\theta_c\right).
\ee
This way, we can check that, on each boundary, the Hamiltonian constraint can be written as the holographic trace flow equation with the $T\bar{T}$ operator
\be
T^i_i=-\frac{1}{2\kappa^2}R_h-\frac{\kappa^2}{2}\left(T_{ij}T^{ij}-(T^i_i)^2\right).
\ee
Note that on the cut-off surface $z=\epsilon$ ($\theta_c=\frac{\pi}{2}$), the energy momentum and its trace vanish.\\
Let us now analyze the Euclidean action on the wedge
\bea\label{on-shell}
I[\theta_c]=-\frac{1}{2\kappa^2}\int_{\mathcal{M}}d^{3}x\sqrt{g}\left(R-2\Lambda\right)-\frac{1}{\kappa^2}\int_{\partial \mathcal{M}}d^2x \sqrt{h}\left(K-1\right).
\eea
We will compactify the $x$ direction by giving it some length $L$, which we send to infinity afterwards. The bulk part of the action evaluated on our region is 
\be
\frac{2 L}{\kappa^2} \int^{\theta_c}_{-\theta_c} \frac{d\theta}{\cos^3\theta}\int^{\infty}_{\frac{\epsilon}{\cos\theta}} \frac{du}{u^2}=\frac{4 L}{\kappa^2\epsilon}\tan \theta_c,
\ee
whereas the boundary contributions are
\be
-\frac{2 L}{\kappa^2}\int^\infty_{\frac{\epsilon}{\cos\theta_c}} du \frac{2\sin\theta_c-1}{u^2\cos^2\theta_c}-\frac{L}{\kappa^2 \epsilon}\int^{\theta_c}_{-\theta_c}  \frac{d\theta}{\cos^2\theta}=-\frac{2L}{\kappa^2 \epsilon}\frac{3\sin\theta_c-1}{\cos\theta_c}.
\ee
Then the on-shell action reads
\be
I[\theta_c]=\frac{2L}{\kappa^2 \epsilon}\frac{1-\sin\theta_c}{\cos\theta_c}.
\ee
By definition
\be
\partial_{\theta_c}\log Z[\theta_c]=-\partial_{\theta_c}I[\theta_c]=\frac{2L}{\kappa^2 \epsilon}\frac{1-\sin\theta_c}{\cos^2\theta_c}.
\ee
On the other hand
\be
\int d^2x\sqrt{h}T^i_i=L\int^\infty_{\frac{\epsilon}{\cos\theta_c}}\frac{du}{u^2}\frac{1}{\cos^2\theta_c}\frac{2}{\kappa^2}\left(1-\sin\theta_c\right)=\frac{2L}{\kappa^2\epsilon}\frac{\left(1-\sin\theta_c\right)}{\cos\theta_c},
\ee
Hence we get
\be
\cos\theta_c\,\partial_{\theta_c}\log Z=\int d^2x\sqrt{h}T^i_i.\label{FLOWT}
\ee
Now, from $\lambda = 2\pi G_N w_c^2$ and \eqref{wteta} our Euclidean time-dependent $T\bar{T}$ coupling is related to $\theta_c$ as
\be
\lambda= 2\pi G_N \left(\frac{1-\sin\theta_c}{\cos\theta_c}\right)^2,
\ee
therefore have
\be
\cos\theta_c\,\partial_{\theta_c}=-2\lambda\partial_\lambda,
\ee
which reproduces the standard relation for the partition function of the $T\bar{T}$ deformed theory in our folded setup.\\
Since our bulk region with regulator at $z=\epsilon$ contains two corners, it may be interesting to study the contribution of the Hayward term at each corner defined as
\be
I_H=-\frac{1}{\kappa^2}\int_{\tilde{\Sigma}}\sqrt{\gamma}\Theta,
\ee
where 
\be
\cos\Theta=n_1\cdot n_2,
\ee
with $n_1$ and $n_2$ being outward-pointing unit-normal vectors to each boundary surface and $\gamma$ being the determinant of the induced metric on the corner $\tilde{\Sigma}$.\\
In our case we have
\be
\cos\Theta=n^\mu_\pm n_\mu=\sin\theta_c.
\ee
with $n^\mu_\pm$ being normal vectors to surfaces at $\theta=\pm \theta$ and $n_\mu$ the normal to the boundary at $z=\epsilon$ or $u=u_0=\frac{\epsilon}{\cos\theta}$. Since $0\le\theta_c\le \pi/2$, $\sin\theta_c=\cos\left(\frac{\pi}{2}-\theta_c\right)$, therefore we have
\be
\Theta=\frac{\pi}{2}-\theta_c.
\ee
The induced metric on the corners at $u=u_0|_{\theta=\pm\theta_c}$ leads to $\sqrt{\gamma}=1/\epsilon$ and the contribution to the action from the Hayward terms on both corners becomes
\be
I_H=2\times\frac{1}{\kappa^2}\int dx \frac{1}{\epsilon}\left(\theta_c-\frac{\pi}{2}\right)=\frac{2 L}{\kappa^2\epsilon}\left(\theta_c-\frac{\pi}{2}\right),
\ee
and the total action including the Hayward term becomes
\be
I[\theta_c]+I_H=\frac{2L}{\kappa^2\epsilon}\left[\frac{1-\sin\theta_c}{\cos\theta_c}+\theta_c-\frac{\pi}{2}\right].
\ee
Then the derivative 
\be
\partial_{\theta_c}\left(I[\theta_c]+I_H\right)=-\frac{2L}{\kappa^2 \epsilon}\left[\frac{1-\sin\theta_c}{\cos^2\theta_c}-1\right],\label{PDI}
\ee
with the last $-1$ in the bracket coming from the Hayward terms.\\
In \cite{Takayanagi:2019tvn}, these corner contributions were argued to correspond to gravity edge-modes. Clearly, they non-trivially modify our trace anomaly as well as the flow equation \eqref{FLOWT}. It will be very interesting to understand Hayward term interpretation from the perspective of $T\bar{T}$ deformations as well as holographic tensor networks discussed here and we hope to return to this problem in the future.

\section{Dimension of Hilbert space in $T\overline{T}$ deformed CFT on interval}\label{sec:dimH}
\begin{figure}[t]
    \centering
    \includegraphics[height=3cm]{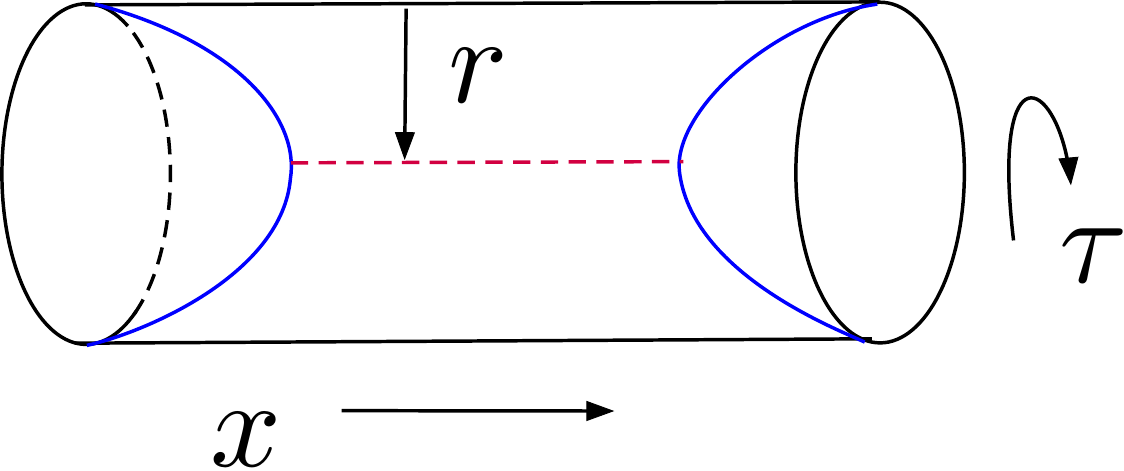}
    \hspace{2cm}
    \includegraphics[height=5cm]{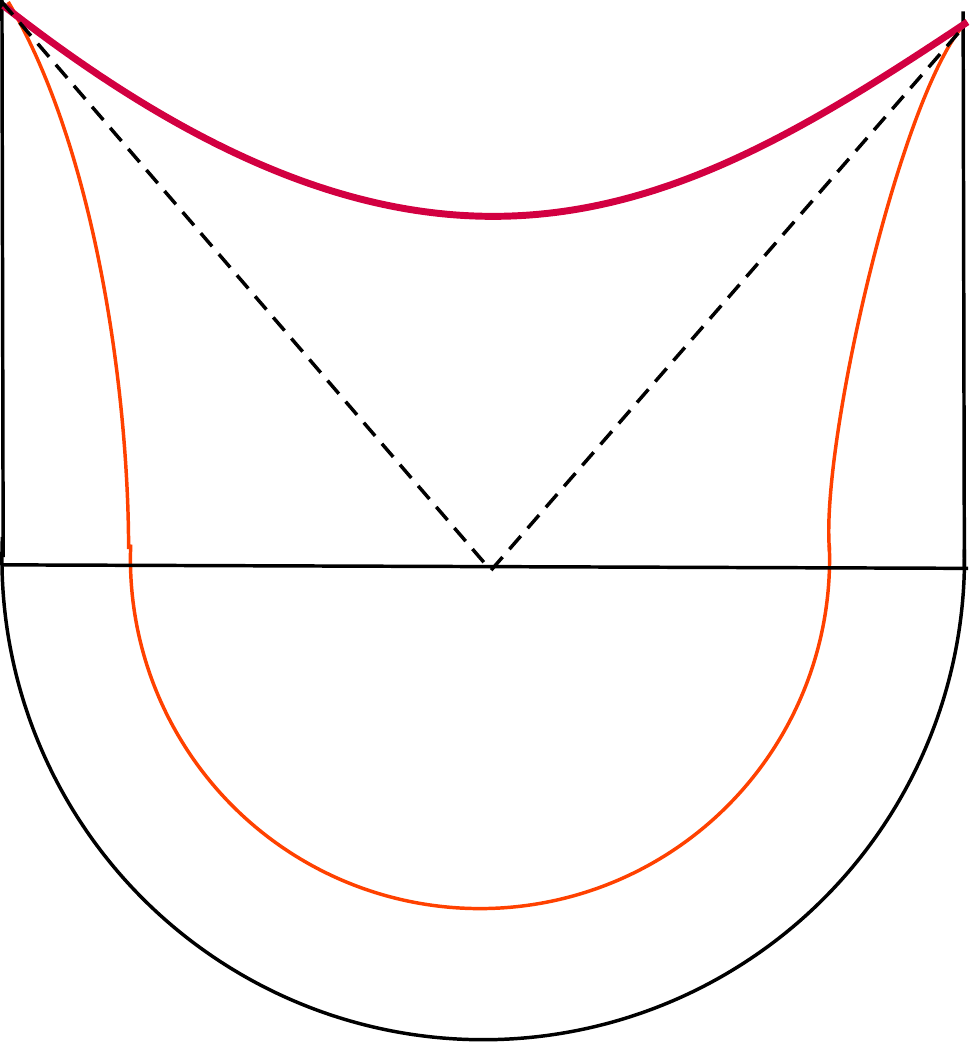}
    \caption{\textbf{Left}: The geometry dual to an interval in the CFT in the small $\beta$ limit. The branes connected to the ends of the interval are shown in blue. The red dotted line denotes the horizon of the black hole. \textbf{Right}: The AdS-Schwarzschild black hole with a cutoff (orange line).}
    \label{fig:brane}
\end{figure}

In this section, we wish to compute the dimension of the Hilbert space of the $T\overline{T}$ deformed CFT on an interval of size $L$ using AdS/CFT. In order to compute the dimension of the Hilbert space, we can study the infinite temperature limit of the partition function on an interval:
\beq
\text{dim}\,\mathcal{H}_{L} = \lim_{\beta \to 0} \text{Tr}\,e^{-\beta H_{\l}},
\eeq
where the subscript $L$ indicates the length of the interval. In order to compute the interval partition function, we need to specify boundary conditions at the ends of the interval. Thankfully, in AdS/CFT, we can implement a natural set of boundary conditions using the AdS/bCFT prescription \cite{Takayanagi:2011zk, Fujita:2011fp}. This amounts to adding a brane in the bulk, with some tension $T$. Since we're working in the $\beta \to 0$ limit, the bulk solution has two (disconnected) branes attached to the two boundaries of the interval (see the left panel of figure \ref{fig:brane}). Furthermore, we are interested in boundary conditions which do not add any additional degrees of freedom at the ends of the interval in the dual CFT. This corresponds to taking $T=0$; for any $T>0$ there is an extra contribution to the entropy coming from the ends of the interval. For $T=0$, the branes just stay at constant $x(r)=x_{\pa}$ in the bulk (where $x_{\pa}$ is the location of the endpoint of the interval in the boundary CFT), and furthermore do not contribute to the on-shell action \cite{Fujita:2011fp}. In this case, therefore, our calculation is equivalent to the torus partition function with the length of the spatial circle being $L$.   

The computation of the torus partition function in AdS/CFT is standard, and involves the Euclidean gravity on-shell action on the AdS-Schwarzschild solution (see the right panel of figure \ref{fig:brane}). The only slight modification here is that we need to work with a finite radial cutoff, since we are interested in the $\ttb$ deformed field theory \cite{McGough:2016lol}. The action is
\beq
S= -\frac{1}{16\pi G_N}\int d^3x\,\sqrt{g}\, (R-2\Lambda) - \frac{1}{8\pi G_N}\int d^2x\,\sqrt{\gamma}\,(K-1),
\eeq
where $\Lambda =-1/l^2$.  We will henceforth set $\ell= 1$. The equation of motion for the metric sets the Ricci scalar to be $R= -6$, thus the on-shell action becomes
\beq
S_{\text{on-shell}} = \frac{1}{4\pi G_N}\int d^3x\,\sqrt{g} - \frac{1}{8\pi G_N}\int d^2x\,\sqrt{\gamma}\,(K-1).
\eeq
The metric of the Euclidean black hole is given by
\beq
g = \frac{dr^2}{h(r)}+ h(r) d\tau^2+ r^2 d\phi^2,\;\; h(r) = r^2 - r_+^2
\eeq
where $r_+ = \frac{2\pi}{\beta_0}$ with $\beta_0$ being the temperature of the CFT dual at asymptotic infinity. If we cut the geometry off at $r=r_c$, then the induced metric on this cut-off slice after scaling out a factor of $r^2$, i.e., $g^{(0)} = r^2 \gamma$, is given by
\beq
\gamma = \left(1-\frac{r_+^2}{r_c^2}\right)d\tau^2 + d\phi^2
\eeq
Thus if we are to interpret the bulk on-shell action in terms of a thermal partition function in the $T\overline{T}$ deformed theory ``living'' at the cutoff surface, then the effective temperature should be identified as 
\beq \label{ET}
\beta = \left(1- \frac{4\pi^2}{\beta_0^2r_c^2}\right)^{1/2}\beta_0.
\eeq
We can compute the on-shell action for this geometry within the cutoff region, and we find\footnote{Note that there is also the other saddle (thermal AdS), which corresponds to the metric
$$
g = \frac{dr^2}{1+r^2}+(1+r^2)d\tau^2+ r^2d\phi^2,
$$
but this does not dominate in the $\beta \to 0$ limit.}
\beqn
S_{\text{on-shell}} &=& \frac{\beta_0 L}{8\pi G_N}\left\{(r_c^2 - r_+^2)+ (- 2r_c^2 + r_+^2) + r_c \sqrt{r_c^2 - r_+^2}\right\},\nonumber\\
&=& \frac{\beta_0 L}{8\pi G_N}\left\{- r_c^2 + r_c \sqrt{r_c^2 - r_+^2}\right\},
\eeqn
where in the first line the first term above comes from the Einstein-Hilbert piece, the second term comes from the Gibbons-Hawking piece, and the last term comes from the counter-term. Thus the partition function becomes
\beq
Z_{grav} = e^{-S_{\text{on-shell}}} = \exp\left\{\frac{ Lr_c}{4 G_N}\left( \frac{r_c}{r_+} -  \sqrt{\frac{r_c^2}{r_+^2} - 1}\right)\right\},
\eeq
where $L$ is the length of the spatial circle, and recall that $r_+ = \frac{2\pi}{\beta_0}$. We can also rewrite this partition function in terms of the effective temperature $\beta$ (defined in \eqref{ET}) by solving for $\beta_0$ in terms of $\beta$ and $r_c$:
\beq \label{PF2}
Z_{grav} = \exp\left\{\frac{ Lr_c}{4 G_N}\left( -\frac{r_c\beta }{2\pi} +  \sqrt{\frac{(r_c\beta)2}{(2\pi)^2} + 1}\right)\right\}.
\eeq
As a sanity check on this formula, note that as $r_c \to \infty$, then the above formula reduces to
\beq
\lim_{r_c\to \infty} Z_{grav}  = e^{\frac{Lr_+}{8G_N}} = e^{\frac{c}{12}\frac{2\pi L}{\beta_0}},
\eeq
where in the second equality we have used $c= 3/2G_N$. This agrees with the universal high-temperature partition function of holographic CFTs. On the other hand, we are interested in the $\beta \to 0$ limit with $r_c$ and $L$ fixed. In this limit, we get
\beq
\lim_{\beta \to 0} Z_{grav}= e^{\frac{Lr_c}{4G_N}},
\eeq
which is notably finite. Thus, we obtain
\beq
\log\,\text{dim}\,\mathcal{H}_L = \frac{Lr_c}{4G_N}=\sqrt{\frac{\pi c}{12\lambda}} L,
\eeq
where we have used the identification $\lambda = \frac{2\pi G_N}{r_c^2}$ of the $\ttb$ coupling. We should mention a caveat here: in order for the $\beta \to 0$ limit to make sense, we should first take the limit $G_N \to 0$ and then send $\beta \to 0$, because otherwise we are not justified in using the classical approximation in the bulk.

\providecommand{\href}[2]{#2}\begingroup\raggedright\endgroup

\end{document}